\documentclass[11pt]{article}
\pdfoutput=1

\usepackage{euscript}
\usepackage{amssymb}
\usepackage{amsfonts}
\usepackage{amsbsy}
\usepackage{epsfig}
\usepackage{amsthm}
\usepackage{amscd}
\usepackage{amstext}
\usepackage{verbatim}
\usepackage{amsmath}
\usepackage{cancel}
\usepackage{capt-of}
\usepackage{empheq}
\usepackage{subfigure}

\usepackage[pdftex]{hyperref}

\def\ben{\begin{equation}}
\def\een{\end{equation}}

\let\a=\alpha \let\b=\beta \let\g=\gamma \let\d=\delta

\let\pa=\partial
\def\be{\begin{equation}}
\def\ee{\end{equation}}
\def\beq{\begin{equation}}
\def\eeq{\end{equation}}
\def\ba{\begin{array}}
\def\ea{\end{array}}

\def\dalemb#1#2{{\vbox{\hrule height .#2pt
       \hbox{\vrule width.#2pt height#1pt \kern#1pt
               \vrule width.#2pt}
       \hrule height.#2pt}}}

\newcommand{\bea}{\begin{eqnarray}}
\newcommand{\eea}{\end{eqnarray}}

\newcommand{\dd}{{\rm d}}

\renewcommand{\eqref}[1]{(\ref{eq:#1})}

\def\ocal{{\mathcal{O}}}

\begin{document}

\begin{center}

{ \Large {\bf
Thermal conductivity at a disordered quantum critical point
}}

\vspace{1cm}

Sean A. Hartnoll$^{1}$, David M. Ramirez$^{1}$  and Jorge E. Santos$^{2}$

\vspace{1cm}

{\small
$^{1}${\it Department of Physics, Stanford University, \\
Stanford, CA 94305-4060, USA }}

\vspace{0.5cm}

{\small
$^{2}${\it Department of Applied Mathematics and Theoretical Physics, \\
University of Cambridge, Wilberforce Road, \\
Cambridge CB3 0WA, UK}}

\vspace{1.6cm}

\end{center}

\begin{abstract}

Strongly disordered and strongly interacting quantum critical points are difficult to 
access with conventional field theoretic methods. They are, however, both experimentally important
and theoretically interesting. In particular,
they are expected to realize universal incoherent transport. Such disordered quantum
critical theories have recently been constructed holographically by deforming a
CFT by marginally relevant disorder. In this paper we find additional disordered fixed points
via relevant disordered deformations of a holographic CFT. Using recently developed methods
in holographic transport,
we characterize the thermal conductivity in both sets of theories in 1+1 dimensions.
The thermal conductivity is found to tend to a constant at low temperatures in one class of fixed points,
and to scale as $T^{0.3}$ in the other. Furthermore, in all cases the thermal conductivity exhibits discrete
scale invariance, with logarithmic in temperature oscillations superimposed on the low temperature scaling behavior.
At no point do we use the replica trick.

\end{abstract}

\pagebreak
\setcounter{page}{1}

\tableofcontents

\newpage

\section{Introduction}

\subsection{Universal incoherent thermal transport}

The thermal conductivity is the simplest and most universally defined quantity in which the subtleties of dc transport arise and
can be studied. In particular, in a translationally invariant system at a nonzero temperature, the
thermal conductivity is generically infinite. This is because at a nonzero temperature there will be a nonzero
energy density and hence an overlap between the thermal current and the momentum operator. Momentum
is conserved by assumption and does not relax. Hence the thermal current also cannot fully relax.

The fate of momentum conservation is especially important in strongly interacting systems.
In a weakly interacting theory, there are many long lived quantities, the quasiparticle number excitations $n_k$.
Momentum is a particular linear combination of the $n_k$ and as such is often not especially privileged. However,
in the absence of quasiparticles, but in the presence of translation invariance, momentum conservation
dominates the late time behavior of heat current excitations.

Strongly interacting transport in the presence of weak momentum relaxation has been extensively studied in
recent years. In particular, the holographic correspondence has provided tractable
explicit models of strongly interacting dynamics. One important upshot of this work has been the development
of the old memory matrix formalism \cite{forster,GW} as the framework of choice to discuss non-quasiparticle transport with weak momentum relaxation \cite{jung, Hartnoll:2007ih, Hartnoll:2008hs, Hartnoll:2012rj, Mahajan:2013cja, Hartnoll:2014gba, Patel:2014jfa, Lucas:2015vna, Lucas:2015pxa}. In particular, if the low energy physics is described by an IR fixed point with an emergent long wavelength translation invariance (as is the case whenever the low energy dynamics admits an effective gapless QFT description), then the memory matrix tells us that the dc thermal conductivity is controlled by the leading irrelevant operator that breaks translation invariance \cite{Hartnoll:2007ih, Hartnoll:2008hs, Hartnoll:2012rj}. Transport in these cases is now solved in principle, up to the characterization of the leading irrelevant operator for a given system.

A more challenging scenario is a strongly interacting theory in which translation invariance is not approximately restored even at the lowest energy scales. In this case the memory matrix is not a useful tool. If the system does not become thermally insulating then the low temperature thermal conductivity is an intrinsic property of the low energy physics. In the context of charge transport, such systems were called `universal incoherent metals' in \cite{Hartnoll:2014lpa}. It was suggested in \cite{Hartnoll:2014lpa} that the cuprates and other `bad metals' should be understood as such universal incoherent metals. The objective of the present work is to provide a controlled theoretical model of universal incoherent thermal transport. The central role played by disorder means that the model is unlikely to be directly relevant to the cuprates, although it may have features in common with superfluid-insulator or quantum hall plateaux transitions. The objective is to gain a theoretical and conceptual handle on incoherent transport. Other recent work towards a theoretical framework for incoherent transport, especially charge transport, can be found in \cite{Lucas:2015lna, Grozdanov:2015qia}.

\subsection{Disordered quantum critical points}

Known instances of gapless low energy theories without translation invariance are disordered quantum critical points (or phases).
These arise when disorder is a relevant deformation of a quantum critical theory \cite{harris, sachdev} which does not gap the theory, but rather the system flows to a new critical point with a finite amount of disorder (yet another possibility are so-called `infinite disorder' fixed points \cite{fisherb,fishera,vojta, sachdev}, these will not be considered here).

In principle, controlled interacting disordered fixed points could be obtained by applying epsilon or large $N$ expansions directly to a replicated field theoretic description of a known CFT, such as the Wilson-Fisher fixed point. However, it is found that the beta function equations do not admit perturbative, stable zeros. Instead, one finds a runaway flow to strong disorder \cite{sachdev, kimwen}. The only known quantum critical disordered fixed points found by this method involve
either continuing to a small number of time dimensions \cite{boyanovsky} or allowing disorder with long range correlation in time \cite{cesare}. It is unclear if these additional approximations can fully capture the quantum disordered dynamics as they are essentially expansions close to the classical statistical physics disordered fixed point \cite{halperin}. In any case, since these fixed points are found in perturbation theory, transport will be described by a quasiparticle-based Boltzmann equation, and will not access the strongly interacting and strongly disordered regimes we are interested in.

Recent work has used the holographic correspondence \cite{Hartnoll:2009sz} to obtain controlled, strongly interacting and strongly disordered fixed points \cite{Hartnoll:2014cua, Hartnoll:2015faa}. Aspects of these fixed points will be recalled below. They were obtained by deforming a CFT by marginally relevant disorder. Past work showed that the (line of) fixed points are characterized by a dynamical critical exponent $z > 1$ that becomes larger as the tunable disorder strength $\bar V$ is increased. In parallel with these developments, major recent work has essentially `solved' the problem of computing d.c.~conductivities in holographic theories \cite{Donos:2014cya, Donos:2014yya, Donos:2015gia, Banks:2015wha}. In particular, the new results give a powerful way to compute the thermal conductivity of complicated disordered spacetimes without needing to explicitly solve perturbation equations about the background. In this work we obtain the thermal conductivity of the holographic disordered fixed points found in \cite{Hartnoll:2014cua, Hartnoll:2015faa}. We will also construct new disordered horizons arising from relevant rather than marginally relevant disorder, and obtain the thermal conductivity of these new fixed points.

\subsection{Results}
\label{sec:resultsA}

The main results in this paper will be for disordered fixed points in 1+1 dimensions. There is no obstruction to considering higher dimensions beyond the fact that more computing power is needed to solve nonlinear partial differential equations in more dimensions. The results for the thermal conductivity are as follows:
\begin{itemize}

\item At weak disorder, in both the marginally relevant and relevant cases, the thermal conductivity $\kappa(T)$ is excellently described by the perturbative formulae (\ref{eq:kappapert}) and (\ref{eq:gammapert}), over the entire temperature range that we can access. See figures \ref{marg001} and \ref{rel001} below.

\item At stronger disorder, the thermal conductivity sees the onset of a low temperature scaling regime. This regime is characterized by a {\bf discrete scale invariance}, so that:
\be\label{eq:kapparesult}
\kappa(T) = T^\alpha F\left( \sin \left[\beta \log \frac{T_0}{T} \right]\right) \,.
\ee
For exponents $\alpha$ and $\beta$, a scale $T_0$ and a function $F(x) \approx c_0 + c_1 x$, for the regimes considered. See figures \ref{margstronger} and \ref{bigrel} below.

\item For the cases of marginally relevant disorder, our results are consistent with the scaling exponent in (\ref{eq:kapparesult}) taking the value $\alpha = 0$, independently of the dynamical critical exponent $z$ in the low energy fixed point theory. The magnitude $c_0$ of the conductivity decreases with increasing strength of disorder. For the relevant disorder, we find $\alpha \approx 0.3$. This latter exponent appears to be independent of the strength of the disorder in the UV and hence a property of a disordered fixed point theory.
\end{itemize}
To our knowledge, these are the first theoretically controlled results on incoherent transport in an explicit strongly disordered and strongly interacting system. In addition
\begin{itemize}
\item For the newly constructed disordered fixed points from relevant disorder we obtain the entropy density as a function of temperature. This thermodynamic quantity is also found to exhibit discrete scale invariance at low temperatures. See figure \ref{entropy}.
\end{itemize}

As anticipated, the thermal conductivity at low temperatures is given by a power of temperature that is intrinsic to the disordered quantum critical low energy theory. These systems therefore indeed exhibit universal incoherent thermal transport. Interestingly, however, the power of temperature that appears  -- $\alpha$ in (\ref{eq:kapparesult}) -- is not the exponent (\ref{eq:scaling}) below that would be anticipated from the simplest dimensional analysis. This is presumably possible because of the scale $T_0$ in the low energy theory (or perhaps, due to a broad disorder distribution at the disordered fixed point), and  suggests that scaling analyses for incoherent systems, such as \cite{Hartnoll:2015sea}, should proceed with caution.

The exponent $\alpha = 0$ in the marginally relevant case indicates that the conductivity takes a finite universal value in these cases (up to oscillations in the logarithm of the temperature). The value depends on where we are in the line of disordered fixed points, but is an intrinsic property of the low energy theory. A universal thermal conductivity is reminiscent of \cite{durst}, where the strength of the disorder appears in two quantities (the finite quasiparticle lifetime and the finite density of states) that cancel in the thermal conductivity. That mechanism, however, is rather perturbative. It seems more likely that the universality here is tied to having $d=1$ spatial dimensions and the fact that the fixed point is obtained as a marginally relevant deformation of a $z=1$ UV theory (from which, perturbatively, a temperature independent thermal conductivity is obtained (\ref{eq:pertmarginal})). By this logic, in $d=2$ spatial dimensions we would expect to obtain $\kappa \sim T$ in the marginal cases. 

Discrete scale invariance has appeared in several previous descriptions of disordered fixed points \cite{d1,d2,d3,boyanovsky, halperin}, in the running of couplings towards the fixed point theory. The appearance here in a radically distinct and
nonperturbative theoretical framework -- and, crucially, without use of the replica trick -- suggests that this is indeed a robust feature of disordered fixed points. Complex scaling exponents in holography are often indicative of dynamical instabilities \cite{Hartnoll:2011fn}, or that the quantum phase transition is pre-empted by a first order transition \cite{Hartnoll:2011pp}. However, the aforementioned perturbative results show that stable spiraling or limit cycle renormalization group flows are possible in disordered systems.

\section{Disordered horizons and disordered quantum criticality}
\label{sec:horizons}

In this section we summarize the holographic description of disordered fixed points \cite{Hartnoll:2014cua, Hartnoll:2015faa}. In doing so, we also recall several general concepts of disordered quantum criticality.

The starting point will be a conformal field theory (CFT) in $d$ spatial dimensions.
This theory is deformed by an operator coupled to a quenched random potential
\be\label{eq:coupling}
{\mathcal L} = {\mathcal L}_\text{CFT} + V(x) \ocal(t,x) \,.
\ee
The operator $\ocal$ has mass scaling dimension $\Delta$ while the disorder is taken to
be short range and Gaussian, so that the disorder averages
\be\label{eq:average}
\langle V(x) \rangle_R = 0 \,, \qquad \langle V(x) V(y) \rangle_R = \bar V^2 \delta^{(d)}(x-y) \,.
\ee
The Harris criterion \cite{harris, sachdev} states that the coupling (\ref{eq:coupling}) is then relevant if
$\Delta < (d+2)/2$.

A minimal holographic setup that can describe a CFT deformed by the random coupling (\ref{eq:coupling})
is gravity coupled to a scalar field, with action
\be\label{eq:action}
S = \frac{1}{16 \pi G_N} \int d^{d+2}x \sqrt{-g} \left(R + \frac{d(d+1)}{L^2} - 2 \left(\nabla \Phi \right)^2
- 4 \, U(\Phi) \right) \,.
\ee
Following the usual holographic dictionary \cite{Hartnoll:2009sz}, a potential of the form $U(\Phi) = - \mu\,\Phi^2/(2 L^2)$ means that the bulk scalar field $\Phi$ is dual to an operator $\ocal$ with dimension $\Delta$ satisfying $\Delta (\Delta - d - 1) = - \mu$. In the works \cite{Hartnoll:2014cua, Hartnoll:2015faa} we set $\mu = d(d+2)/4$, so that $\Delta = (d+2)/2$ and the disorder is marginal, saturating the Harris criterion. In particular, this means that the disorder strength $\bar V$ in (\ref{eq:average}) is a dimensionless free parameter. In section \ref{sec:relevant} of this paper we will also consider the case of relevant disorder with $\Delta = 3/4$ in $d=1$ dimensions (for context, marginal disorder has $\Delta = 3/2$ in $d=1$, and the unitarity boundary is $\Delta = 0$).

The papers \cite{Hartnoll:2014cua, Hartnoll:2015faa} asked: what is the low energy dynamics of the deformed theory (\ref{eq:coupling})? In holography this question is studied \cite{Hartnoll:2009sz} by solving the Einstein equations of motion following from the action (\ref{eq:action}) subject to the boundary condition that the scalar field tend to the disorder potential $V$ at the asymptotic boundary of the spacetime: $\lim_{r \to 0} \left[r^{\Delta-d-1} \Phi(r,x)\right] = V(x)$. The Einstein equations with a disordered source are complicated. They can be solved numerically or perturbatively in weak disorder. With the solution at hand, the low energy physics -- such as entropy density at low temperatures --  is characterized by the properties of the spacetime deep in the interior, where the holographic coordinate $r \to \infty$. The results of \cite{Hartnoll:2014cua, Hartnoll:2015faa} will be reviewed below. Other works constructing disordered holographic spacetimes include \cite{Adams:2011rj, Adams:2012yi, Arean:2013mta, Zeng:2013yoa, Arean:2014oaa, Arean:2015sqa}.

The disordered potential $V(x)$ is modeled by a sum over $N$ cosines with increasing wavevector and random phases. 
We will not give details here, see \cite{Hartnoll:2014cua, Hartnoll:2015faa}.\footnote{A small difference in notation compared to the earlier papers: in this paper we use $d$ to denote the number of boundary spatial dimensions. Also, we have corrected a factor of $2\pi$ in the normalization of $\bar V$ in \cite{Hartnoll:2014cua, Hartnoll:2015faa}, so that now $\bar V^2 = \bar V^2_\text{correct} = 2 \pi \bar V^2_\text{old}$. To facilitate comparison with previous works, especially for numerical results, we will sometimes quote the result in terms of $\hat V = \bar V_\text{old}$, so that $\bar V^2 \equiv 2 \pi \hat V^2$.} There is both a short and long distance cutoff on the wavevectors, that we will denote as $k_\text{UV}$ and $k_\text{IR}$ respectively. Note that, $k_\text{IR} = k_\text{UV}/N$. This translates into a temperature range of roughly $k_\text{IR} \lesssim T \lesssim k_\text{UV}$ over which the system is truly disordered. Thus, accessing disordered physics at low temperatures requires a large number $N$ of oscillatory modes. In fact, as explained in \cite{Hartnoll:2015faa}, one is able to go to slightly lower temperatures than the rough estimate suggests thanks to favorable numerical factors.

\section{Perturbative formula for the thermal conductivity}

In a regime where the disorder may be treated perturbatively, analytic field theoretic results are possible
using the memory matrix formalism. For a recent clean discussion of the hydrodynamics of a
strongly interacting CFT perturbed by weak disorder see \cite{Davison:2014lua}.
The thermal conductivity is
\be\label{eq:kappapert}
\kappa_\text{pert.} = \frac{s}{\Gamma} \,,
\ee
where $s$ is the entropy density, and the momentum relaxation rate
\cite{Hartnoll:2007ih, Hartnoll:2008hs, Hartnoll:2012rj}
\be\label{eq:gamma}
\Gamma = \frac{\bar V^2}{s T} \lim_{\omega \to 0} \int \frac{d^dk}{(2\pi)^d} \frac{k^2}{d} \frac{\text{Im} \, G^R_{\ocal\ocal}(\omega,k)}{\omega} \,.
\ee
Here $d$ is the number of spatial dimensions and $G^R_{\ocal\ocal}(\omega,k)$ is the retarded Green's function
of the operator that couples to disorder in (\ref{eq:coupling}). For the case of strongly interacting theories
without an underlying relativistic scale invariance see \cite{Mahajan:2013cja, Lucas:2015pxa}.

From the above two formulae one can immediately obtain the temperature scaling of the disorder at temperatures
$k_\text{IR} \lesssim T \lesssim k_\text{UV}$ (this is roughly the range of temperatures that is insensitive to the cutoffs on the disorder distribution). Simple power counting gives \cite{Hartnoll:2007ih, Hartnoll:2008hs}
\be\label{eq:pertscaling}
\kappa_\text{pert.} \sim \frac{T^{2d+1 - 2 \Delta}}{\bar V^2} \,.
\ee
Here we used the fact that $s\sim T^d$ for a CFT. In particular, for the marginal case
\be\label{eq:pertmarginal}
\kappa_\text{pert.} \sim \frac{T^{d-1}}{\bar V^2}  \,.
\ee
These expressions use the scaling symmetries of the high energy CFT fixed point.
They need not hold at the lowest temperatures, even for small $\bar V$, as there
the system has been driven to the disordered IR fixed point by the relevant or marginally
relevant deformation (\ref{eq:coupling}). Understanding the thermal conductivity in the
far IR is the primary objective of this paper, and we will return to that question shortly.
At weak disorder, the perturbative thermal conductivity (\ref{eq:pertscaling})
can be expected to appear at intermediate temperatures.
Reproducing the perturbative answer will build confidence in our more general results.

Specializing to the holographic theory allows us to compute
the retarded Green's function $G^R_{\ocal\ocal}(\omega,k)$ in (\ref{eq:gamma}) explicitly.
In this computation and in the remainder of the paper we will focus on one boundary space dimension ($d=1$). Both the holographic formula for dc conductivities and the numerics we shall
employ below are substantially simpler in this case. There is no fundamental obstacle, beyond computing
power, to study of the higher dimensional case. The details of the (somewhat standard) holographic computation of $G^R_{\ocal\ocal}(\omega,k)$
are given in Appendix \ref{sec:Green}. The perturbative momentum relaxation rate is found to be
\be\label{eq:gammapert}
\Gamma = \frac{\bar V^2 L}{2 \pi G_N} \frac{1}{sT} (2 \pi T)^{2 \Delta - 3} \frac{\sin^2(\pi \Delta)}{\pi^2} \Gamma(2-\Delta)^2 \int_{k_\text{IR}}^{k_\text{UV}} \frac{dk}{2\pi} k^2 \left| \Gamma \left(\frac{\Delta}{2} + \frac{i k}{4 \pi T} \right) \right|^{4} \,.
\ee
We have explicitly included the long and short wavelength cutoffs on the disorder in order for better comparison with the numerical results below. The perturbative thermal conductivity (\ref{eq:kappapert}) is then found by recalling that for the three dimensional Schwarzschild-AdS black hole that we are perturbing about, the entropy density $s = (\pi L)/(2 G_N) \times T$.

\section{Holographic formula for the thermal conductivity}

In a sequence of beautiful recent works, Donos and Gauntlett have obtained a general nonperturbative formula for the thermal and electric dc conductivities of holographic theories \cite{Donos:2014cya, Donos:2014yya, Donos:2015gia, Banks:2015wha}. Their expressions are a major generalization of the older important result by Iqbal and Liu \cite{Iqbal:2008by}. Specifically, for a class of holographic theories, they have derived an expression for the dc conductivities in terms of data evaluated purely on the event horizon. Crucially for our present purposes, the formula holds with arbitrary spatially dependent sources in the dual quantum field theory. These results allow us to go beyond perturbation theory in the disorder.

In this section we adapt the results of \cite{Donos:2014yya} to the case of the Einstein-scalar theory (\ref{eq:action}) in three bulk spacetime dimensions. Scalar fields have also recently been considered in \cite{Rangamani:2015hka, Banks:2015wha}. The details of the computation can be found in Appendix \ref{sec:horizon}. The full spacetime metric is written as
\be\label{eq:metric}
ds^2 =  \frac{L^2}{y^2} \left[y_+^2 \left( -f(y) A(y,x) dt^2 + S(y,x) \left(dx + F(y,x) dy \right)^2 \right) + \frac{B(y,x)}{f(y)} dy^2 \right] \,.
\ee
Here $f(y) = 1 - y^2$. This form of the metric will be useful for the numerics in the following section.
In these coordinates, the horizon is at $y=1$ and the asymptotic boundary at $y=0$.
At the boundary we impose $A = B = S = 1$ and $F=0$, so that the only source will be in the scalar field $\Phi \sim r^{d+1-\Delta} V(x)$. At the horizon $A = B = A^{(0)}(x)$, $S = S^{(0)}(x)$, the scalar $\Phi = \Phi^{(0)}(x)$, and $F$ again vanishes.
With these boundary conditions, the constant $y_+$ determines the temperature as $y_+ = 2 \pi T$.
In Appendix \ref{sec:horizon} the thermal conductivity is found to be
\be\label{eq:kappa}
\kappa = \frac{\pi^2 T^2}{2 G_N} \left[ \frac{1}{L_x} \int dx \frac{\left(\pa_x \Phi^{(0)}\right)^2}{L \sqrt{S^{(0)}}} \right]^{-1} \,.
\ee
Here $L_x$ is the length of the $x$ direction in the dual field theory.

To evaluate the integral in (\ref{eq:kappa}), we must solve the Einstein-scalar equations of motion numerically in order to find the disordered black hole background and hence $\Phi^{(0)}(x)$ and $S^{(0)}(x)$. The virtue of (\ref{eq:kappa}) is that we do not
need to additionally solve numerical perturbation equations about the background. In Appendix \ref{sec:horizon} we check that
(\ref{eq:kappa}) reproduces the perturbative memory matrix result (\ref{eq:kappapert}) at weak disorder. Schematically: To leading order in small $\bar V$, $S^{(0)} = 1$ and $\Phi$ describes a linearized perturbation on top of Schwarzschild-$AdS_3$. The integral 
$\int dx \left(\pa_x \Phi^{(0)}\right)^2$ appearing in (\ref{eq:kappa}) can then be written in terms of the low frequency limit of the retarded Green's function of $\ocal$ appearing in (\ref{eq:kappapert}). In fact, it is of note that the full holographic expression (\ref{eq:kappa}) looks like a local version of the perturbative result (\ref{eq:kappapert}), in which $S^{(0)}$ plays the role of a spatially varying entropy density. This `hydrostatic' aspect of holographic dc conductivities has recently been emphasized in \cite{Lucas:2015lna}.

\section{Numerical results}
\label{sec:results}

Consider first the case of marginal disorder: $\Delta = 3/2$ in $d=1$. For this case, we have constructed the numerical solutions using the same methods as we described in \cite{Hartnoll:2015faa}. The spacetime takes the form (\ref{eq:metric}). We solve the Einstein-scalar equations following from the action (\ref{eq:action}), subject to the boundary conditions described below equation (\ref{eq:metric}) above. From the solution, the thermal conductivity is obtained from horizon data using (\ref{eq:kappa}). All numerics in this section are performed with a sum over $N=50$ oscillatory modes in the disorder potential $V(x)$.

\subsection{Thermodynamics and scaling expectation for the conductivity}

It is found that in the CFT deformed by marginal disorder, the disorder grows logarithmically towards low energies (i.e. towards the interior of the spacetime), and is therefore marginally relevant \cite{Hartnoll:2014cua, Hartnoll:2015faa}. Such logarithms were first found holographically in \cite{Adams:2012yi}.

A basic quantity characterizing any disordered fixed point is the dynamical critical exponent $z$. In \cite{Hartnoll:2014cua} it was found that the disorder averaged spacetime metric in the far interior, at zero temperature, takes a `Lifshitz' \cite{Kachru:2008yh} scaling form
\be\label{eq:IRmetric}
\lim_{r \to \infty} \langle ds^2 \rangle_R \sim- \frac{dt^2}{r^{2z}} + \frac{dr^2 + d\vec x^2}{r^2}  \,.
\ee
The dynamic critical exponent $z$ was determined analytically in a perturbation theory in $\bar V$ and numerically in general. Furthermore, heating up by a small temperature $T$, the entropy density was shown to scale as
\be
s \sim T^{d/z} \,,
\ee
with the same $z$ as appeared in the disorder averaged metric (\ref{eq:IRmetric}).\footnote{It should be noted that the difference between the scaling form $s \sim T^{1/z}$ (in $d=1$) and the perturbative result $s \sim T\left(1 + \left[\frac{1}{z}-1\right] \log T + \cdots \right)$ can be quite small when $z \approx 1$. In particular, in perturbation theory, the logarithmic derivative $\frac{T}{s} \frac{ds}{dT} = \frac{1}{z} + \left(\frac{1}{z}-1\right)^2 \log T + \cdots $ is close to the exponentiated value of $\frac{1}{z}$, even at low $T$. However, at $\hat V = 1$ one has $z \approx 2$ \cite{Hartnoll:2014cua}, and the hence the constant logarithmic derivative found in \cite{Hartnoll:2015faa} requires the exponentiated scaling behavior of the entropy density.}
 These last two results above demonstrate the emergence of a disordered quantum critical point at low energies, with an emergent scale invariance characterized by the exponent $z$. The fact that $z > 1$ indicates that disorder is playing a finite role in the fixed point physics.

Before turning to numerical results for the thermal conductivity, the thermodynamic facts above suggest the following simple scaling analysis. The entropy scaling $s \sim T^{d/z}$ suggests that hyperscaling is obeyed and that there are no scales remaining at the disordered IR fixed point. In particular, this scaling suggests that the energy density $\epsilon$ has its canonical dimension $[\epsilon] = d + z$. The conservation equation $\dot \epsilon + \nabla \cdot j^Q = 0$ then determines the dimension of the heat current to be $[j^Q] = d-1 + 2 z $. The thermal conductivity is defined via $j^Q = {-} \kappa \nabla T$, and hence $[\kappa] = d + z - 2$. We conclude that
  \begin{align}\label{eq:scaling}
    \kappa_\text{scaling} \sim T^{(d+z-2)/z} \xrightarrow[d=1]{} T^{1 - 1/z}\, .
  \end{align}
This result gives $\kappa \sim T^0$ when $z=1$, consistent with the perturbatively marginal disorder expression (\ref{eq:pertmarginal}) when $d=1$. However, for any arbitrarily small disorder strength we have just recalled that the IR fixed point has $z > 1$. The scaling result (\ref{eq:scaling}) would therefore suggest that the thermal conductivity should go to zero like (\ref{eq:scaling}) at the lowest temperatures, as the marginally relevant disorder drives the system away from the UV fixed point. This expectation will be seen to fail.

\subsection{Thermal conductivity}

\begin{figure}[h]
\centering
\includegraphics[height = 0.3\textheight]{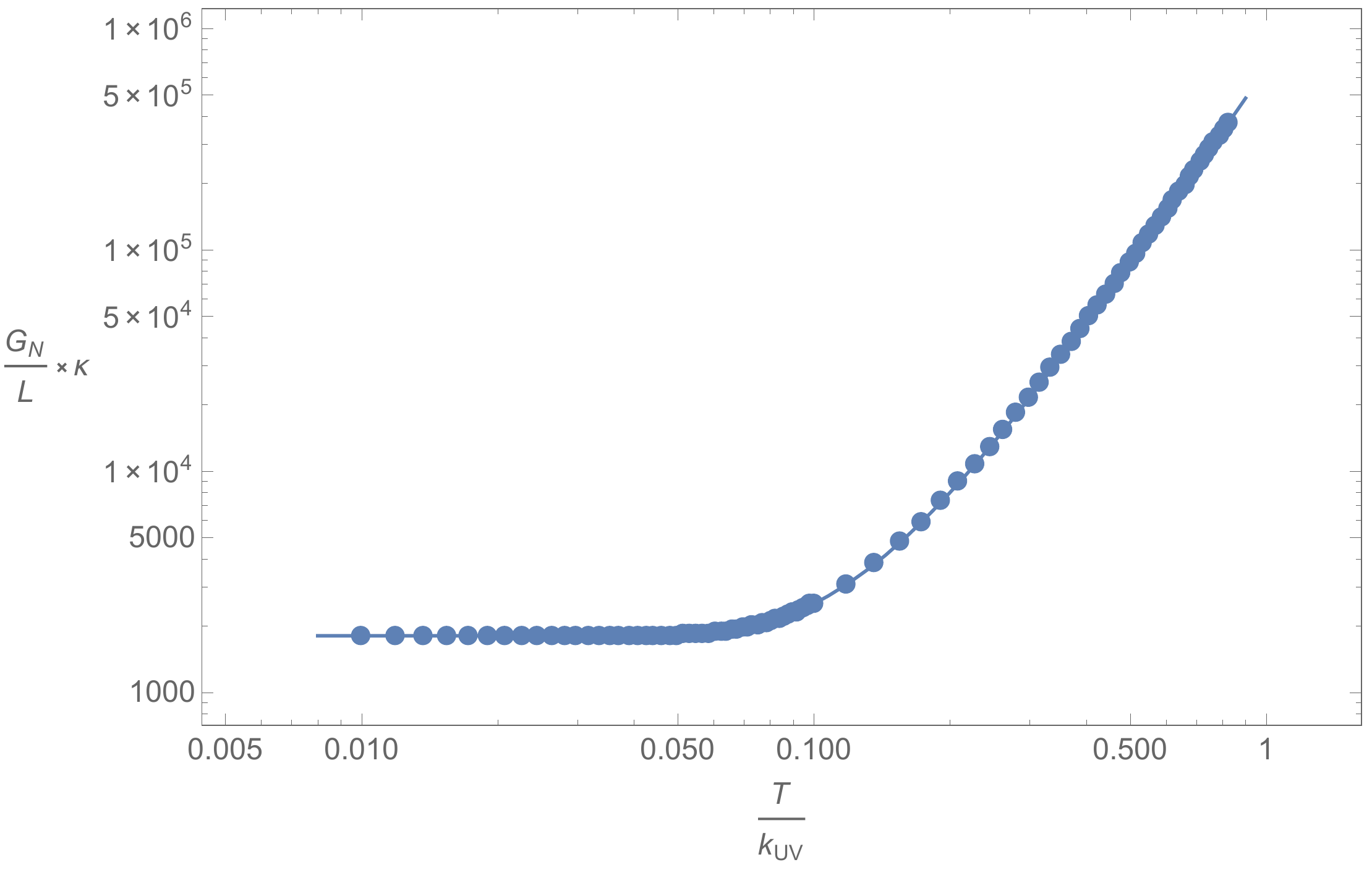}
\caption{\label{marg001}  {\bf Log-log plot of thermal conductivity $\kappa(T)$} with $\hat V \equiv \bar V/\sqrt{2 \pi} = 0.01$ and marginal disorder ($\Delta = \frac{3}{2}, d=1$). The dots are numerical data points with $N=50$ oscillator modes generating the disordered potential. The solid line is the analytic perturbative result (\ref{eq:kappapert}) and (\ref{eq:gammapert}). At low temperatures $\kappa \sim T^0$. At high temperatures compared to the disorder cutoff $k_\text{UV}$, $\kappa \sim T^3$.}
\end{figure}

Figure \ref{marg001} shows the thermal conductivity as a function of temperature for weak disorder, $\hat V \equiv \bar V/\sqrt{2 \pi} = 0.01$. Also shown in the same plot is the perturbative result (\ref{eq:kappapert}) and (\ref{eq:gammapert}). The numerical and perturbative results are seen to be in excellent agreement over the entire temperature range. We have made a log log plot to emphasize the two different power law regimes. The high temperature power law asymptotes to $\kappa \sim T^3$ at temperatures $T \gtrsim k_\text{UV}$, i.e. above the short distance cutoff on the disorder distribution, as can be seen from the perturbative result (\ref{eq:gammapert}). In the truly disordered regime, the thermal conductivity is seen to be a constant. This regime extends over $0.01 \lesssim T/k_\text{UV} \lesssim 0.06$ in the plot. There is no sign over this temperature range of the IR cutoff on the disorder distribution. While $k_\text{IR} = k_\text{UV}/N = 0.02 \, k_\text{UV}$ here, numerical factors shift the effect of the cutoffs to lower temperatures \cite{Hartnoll:2015faa}.

\begin{figure}[h]
\centering
\subfigure{\includegraphics[height = 0.2\textheight]{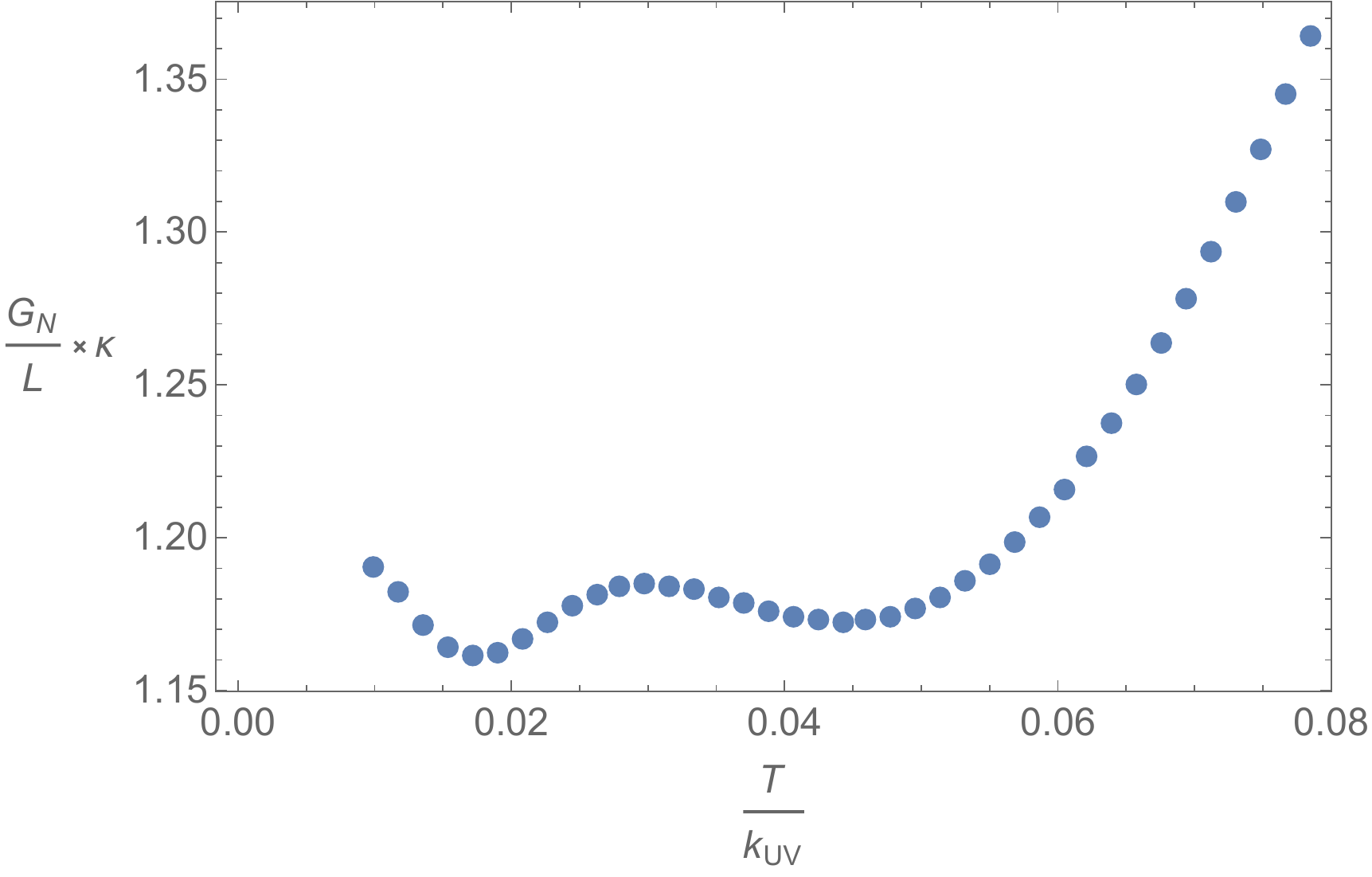}}
\subfigure{\includegraphics[height = 0.2\textheight]{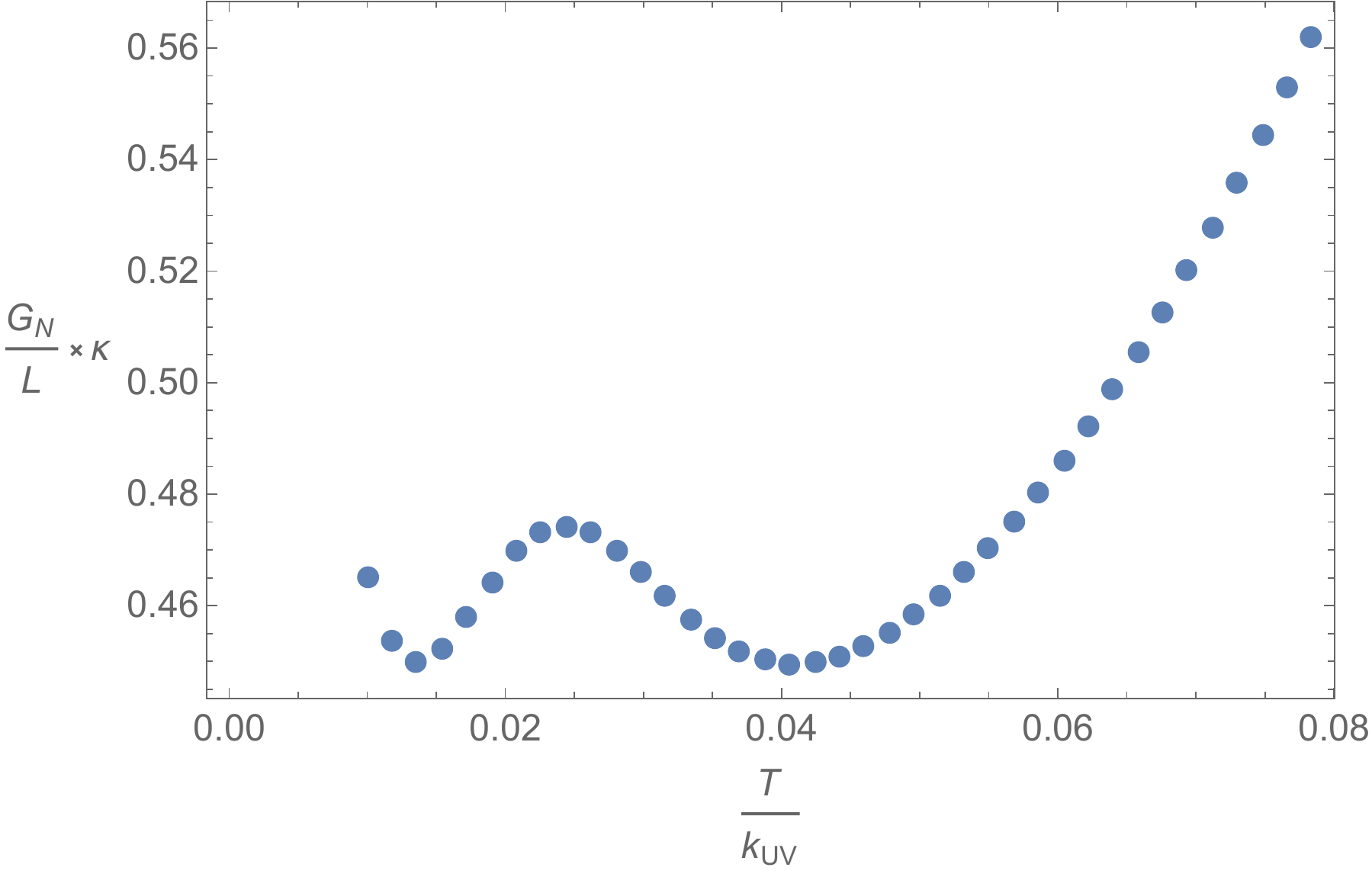}}
\caption{\label{margstronger} {\bf Low temperature thermal conductivity $\kappa(T)$} for marginal disorder with $\hat V = 0.5$ (left) and $\hat V = 1$ (right). The disorder has been simulated with $N=50$ oscillatory modes. The perturbative formulae (\ref{eq:kappapert}) and (\ref{eq:gammapert}) predict the zero temperature values of the conductivity to be $0.72$ and $0.18$, respectively, and therefore underestimate the true result.}
\end{figure}

Bolstered by the agreement of the previous figure, we can now turn to stronger disorder. Figure \ref{margstronger} shows the thermal conductivity as a function of temperature for $\hat V = 0.5$ and $\hat V = 1$. The plots focus on the low temperature regime, away from the short and long distance cutoffs on the disorder.
These are the temperatures over which the scaling of the entropy density $s \sim T^{1/z}$ was found in \cite{Hartnoll:2015faa}. Once again, the thermal conductivity saturates to a constant value at low temperatures. The value of the low temperature conductivity is no longer close to the perturbative prediction. A new phenomenon appears, furthermore, which is that the thermal conductivity oscillates at low temperatures.

In the following section we will consider relevant disorder. For the relevant disorder we will be able to go to lower temperatures and see more oscillations than are visible in figure \ref{margstronger}. The perseverance of the oscillations with stronger disorder suggests that the oscillations continue down towards zero temperature and are not merely an intermediate scale phenomenon. The simplest explanation is that they correspond to an imaginary scaling exponent for the thermal conductivity, which leads to log-periodic oscillations (discrete scale invariance) of the form advertised in equation (\ref{eq:kapparesult}) above. We mentioned previously in section \ref{sec:resultsA} above that discrete scale invariance has appeared in previous perturbative field theoretic studies of disordered fixed points, but had been suspected to be an artifact of the replica trick. We are not using any replica trick here. We will do a detailed fitting to the parameters in (\ref{eq:kapparesult}) in the following section on relevant disorder, where we are able to get to lower temperatures.

The fact that the conductivity tends to a constant (up to oscillations) at low temperatures is in disagreement with the simplest scaling expectation (\ref{eq:scaling}). For instance, equation (\ref{eq:scaling}) would predict $\kappa \sim T^{0.5}$ for the $\hat V = 1$ case, for which $z \approx 2$ \cite{Hartnoll:2014cua, Hartnoll:2015faa}. This is not seen in figure \ref{margstronger}. Instead the results in figure \ref{margstronger} suggest that the low temperature scaling of $\kappa \sim T^0$ is correctly predicted by the perturbative formula (\ref{eq:pertmarginal}), up to the appearance of oscillations and a renormalization of the overall value of the conductivity. At present we do not understand the physics at work here.

\section{Relevant disorder}
\label{sec:relevant}

This section gives evidence for a new class of disordered fixed points, obtained by deforming a CFT by relevant rather than marginal disorder. Because these are new gravitational solutions, we will describe how they have been constructed and characterize their thermodynamics before going on to present the thermal conductivity.

\subsection{Numerical methods}

For sufficiently negative mass squared, there are two choices of quantization in asymptotically Anti-de Sitter spacetime. Motivated by the expectation that disorder effects should be stronger for the choices of $\Delta$ that most violate the Harris criterion, we decided to work in alternative quantization. For reasons we explain shortly, we chose $\mu = 15/16$ in the action (\ref{eq:action}) which, using alternative quantization, leads to $\Delta = 3/4$. This in turn means that, in Fefferman-Graham coordinates, the scalar field approaches the boundary ($z \to 0$) as
\begin{equation}
\Phi(z,x) = z^{3/4} \left[\langle\mathcal{O}(x)\rangle + \sqrt{z}\,V(x)+\cdots\right]\,,
\label{eq:expansion}
\end{equation}
where $V(x)$ is the disorder potential.

We will be using spectral collocation methods which, in their simplest formulation, are best tailored for solving PDEs where all variables are analytic in their domain of integration. This suggests the line element (\ref{eq:metric}) should be changed in order to accommodate the slow decay exhibited by the scalar field in (\ref{eq:expansion}). We choose:
\be
ds^2 =  \frac{L^2}{\tilde{y}^4} \left[y_+^2 \left( -g(\tilde{y}) A(\tilde{y},x) dt^2 + S(\tilde{y},x) \left(dx + 2\,\tilde{y}\,F(\tilde{y},x) d\tilde{y} \right)^2 \right) + \frac{4\tilde{y}^2 B(\tilde{y},x)}{g(\tilde{y})} d\tilde{y}^2 \right] \,,
\label{eq:metricalternative}
\ee
where $g(\tilde{y})=1-\tilde{y}^4$. If we set $A=B=S=1$ and $F=0$, we recover the BTZ black hole. For the scalar field we set $\Phi(\tilde{y},x) = \tilde{y}^{3/2}\,\varphi(\tilde{y},x)$. In order to solve the resulting system of PDEs, we follow \emph{mutatis mutandis} \cite{Hartnoll:2015faa}. In particular, we use the De-Turck trick \cite{Headrick:2009pv}, with the BTZ black hole being the reference metric. The only change worth mentioning is the boundary condition for the scalar field $\Phi$ (or alternatively, for $\varphi$). We first note that, asymptotically, the relation between $\tilde{y}$ and the usual Fefferman-Graham coordinates is simply $\tilde{y} = \sqrt{y_+} \sqrt{z}[1+\mathcal{O}(z^{3/2})]$. This implies that if we want to recover the expansion (\ref{eq:expansion}) we need to set a Neumann-type boundary condition for $\varphi$ of the form
\begin{equation}
\left.\partial_{\tilde{y}} \varphi(\tilde{y},x)\right|_{\tilde{y}=0} = y_{+}^{-5/4} V(x)\,.
\end{equation}
The variables are analytic in these new coordinates. The fact that analyticity can be achieved through a relatively simple change of coordinates is the reason we focus on $\Delta = 3/4$.

\begin{figure}[h]
\centering
\subfigure{\includegraphics[height = 0.3\textheight]{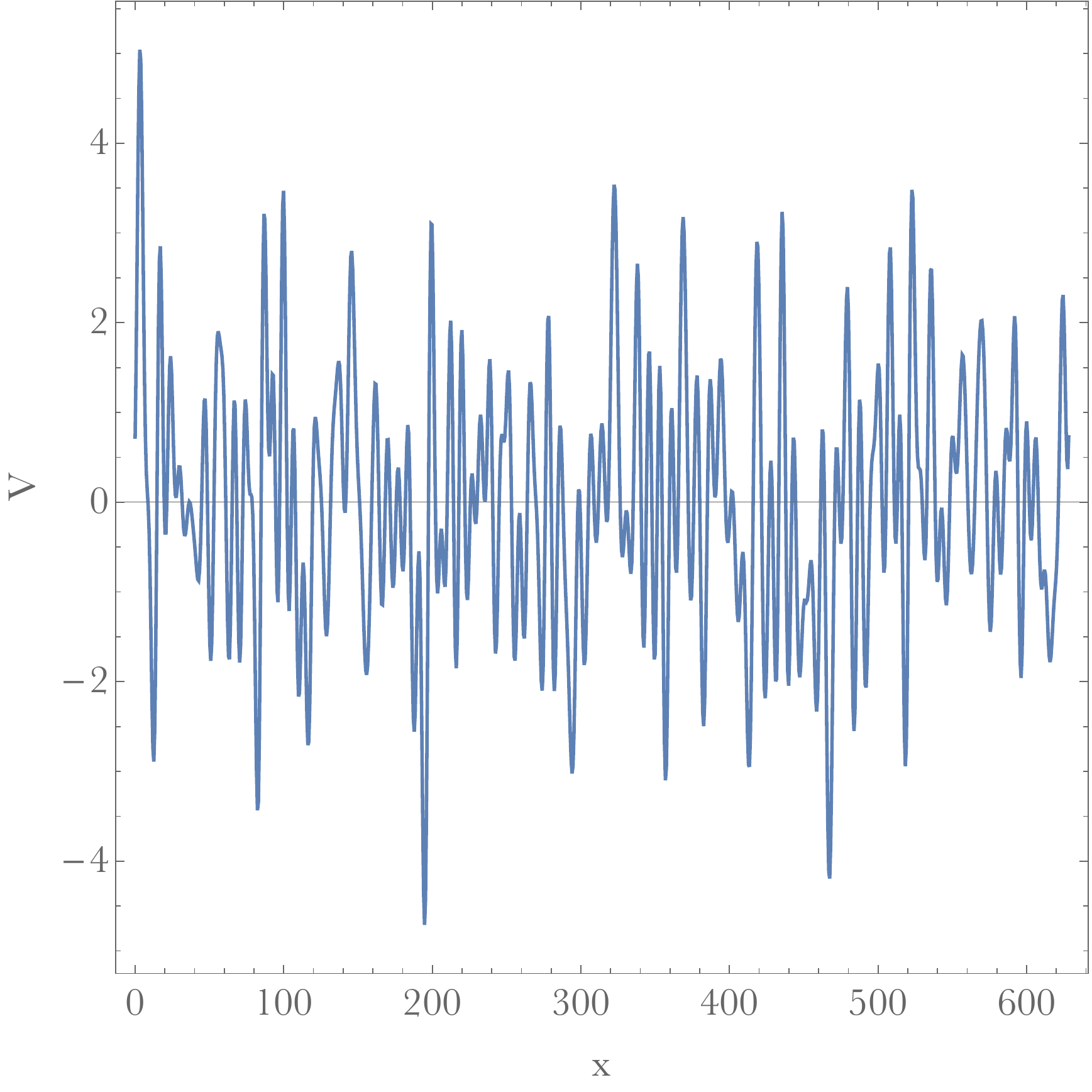}}
\subfigure{\includegraphics[height = 0.3\textheight]{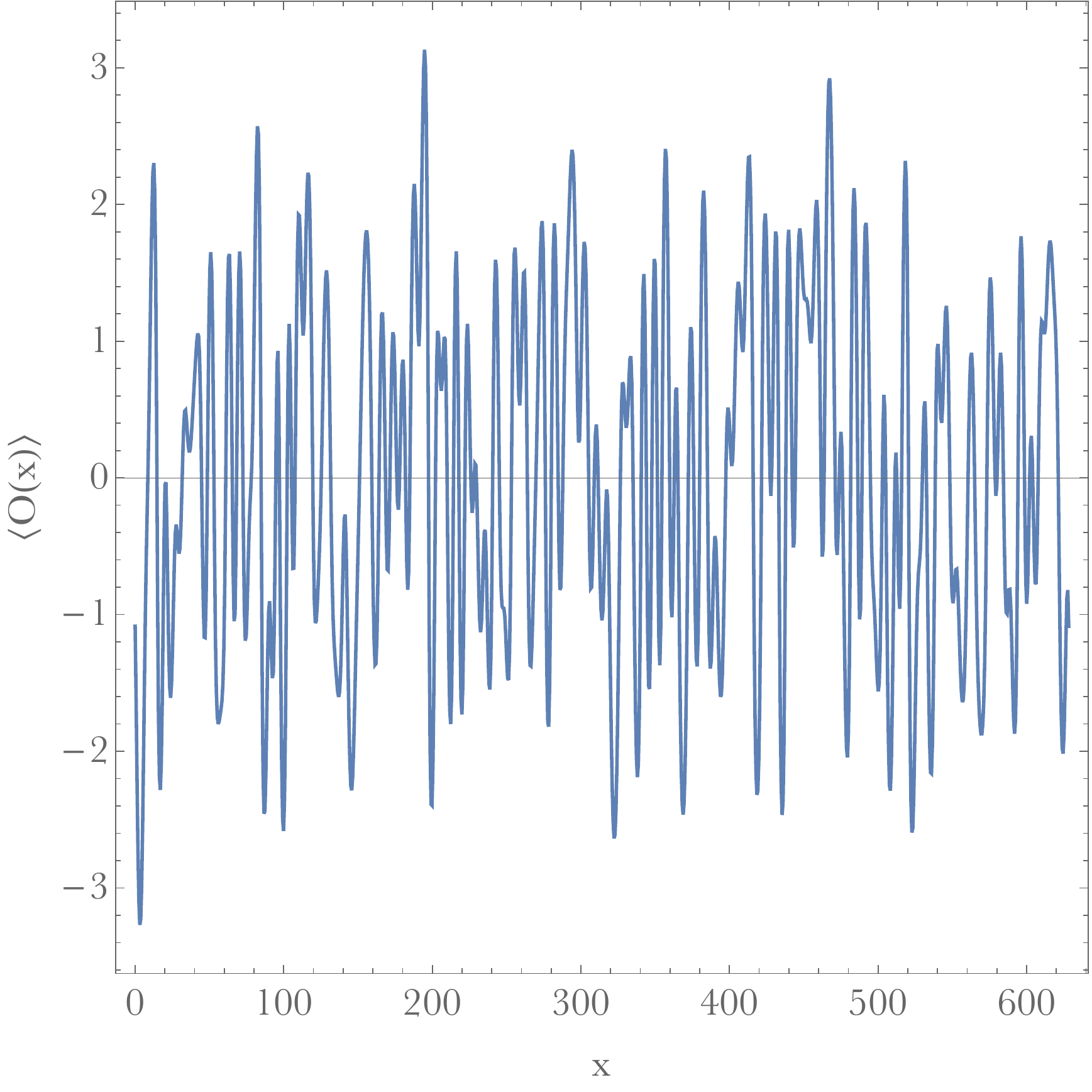}}\\
\subfigure{\includegraphics[height = 0.3\textheight]{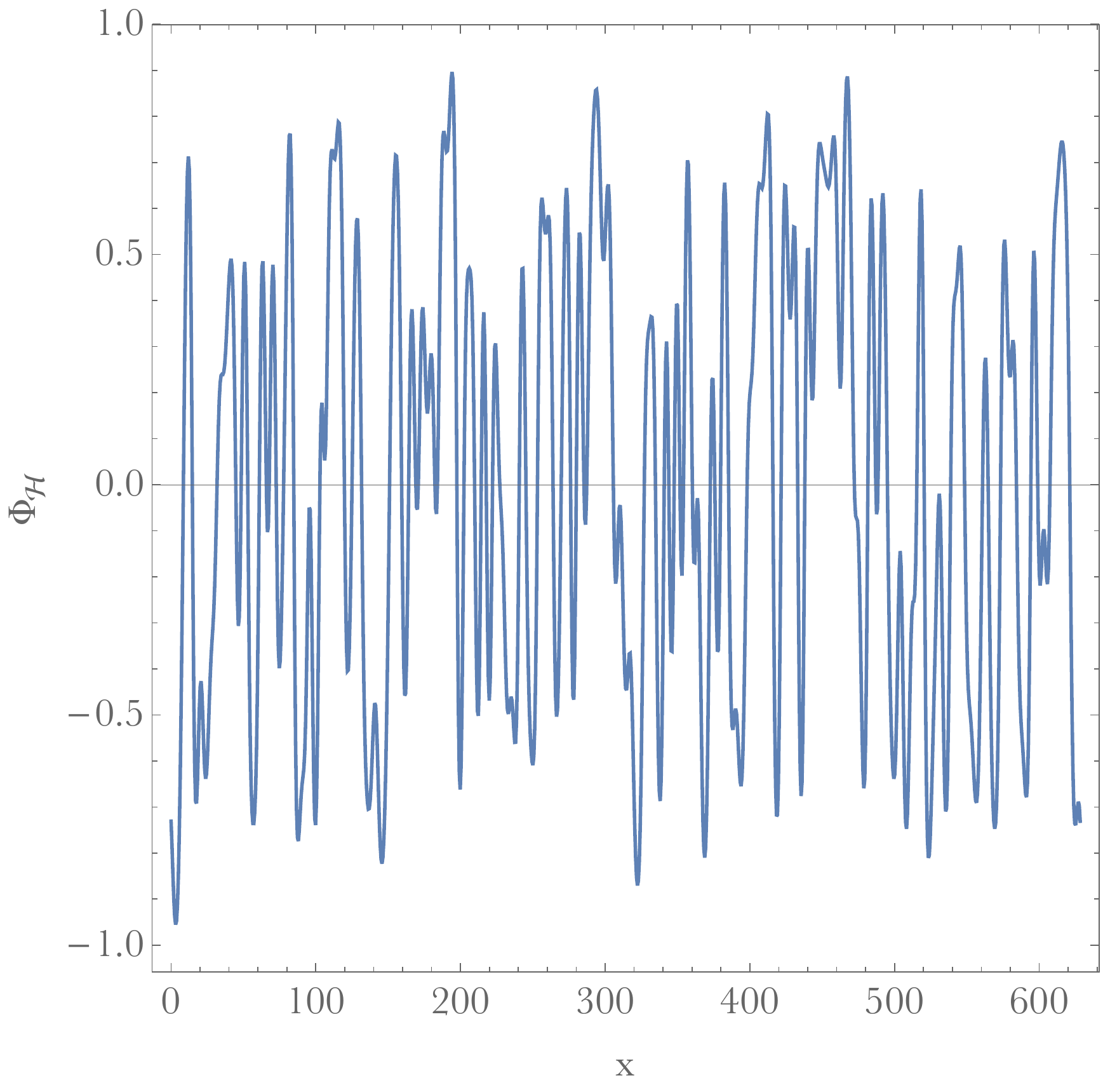}}
\subfigure{\includegraphics[height = 0.3\textheight]{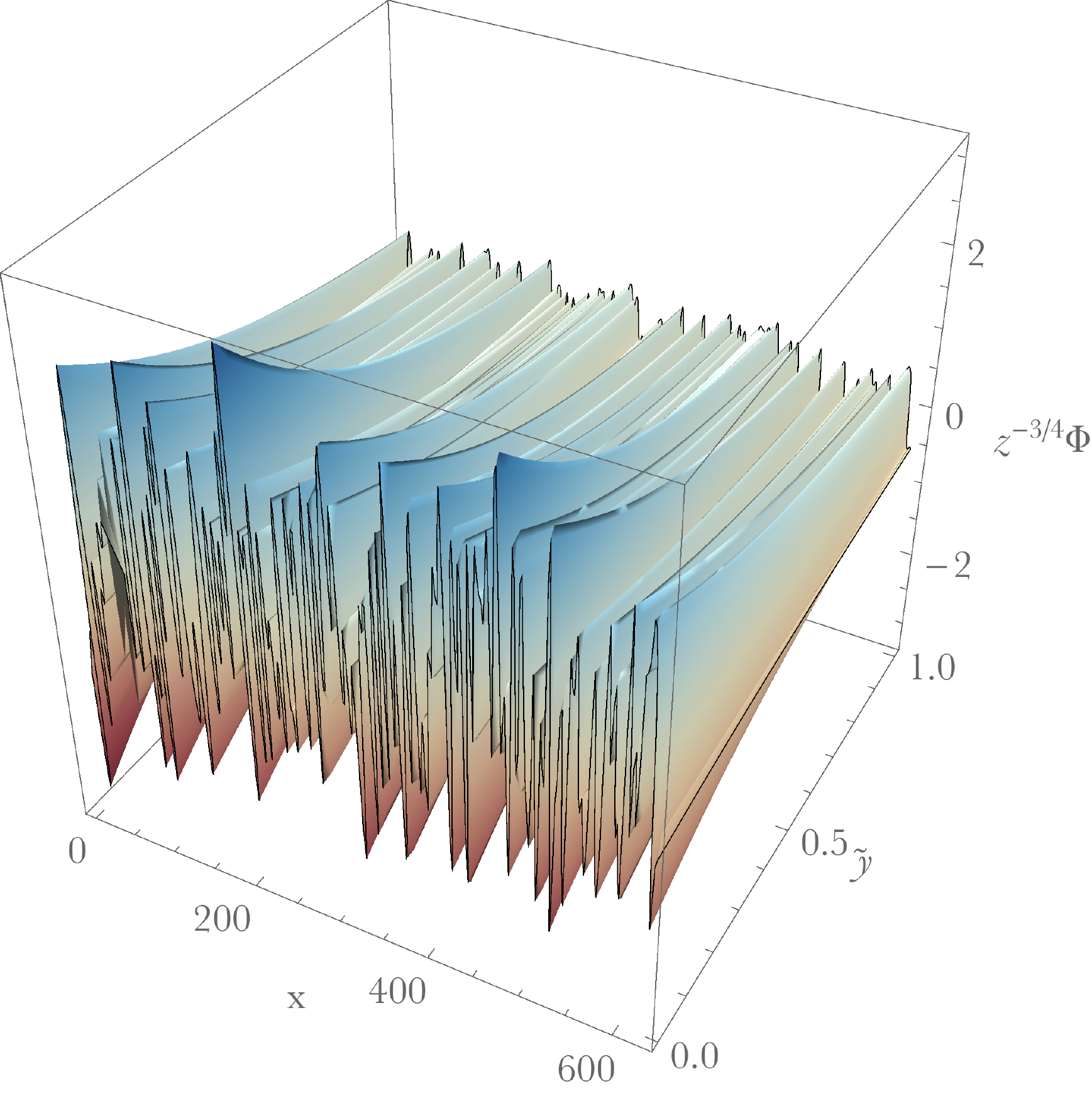}}
\caption{\label{numerics} {\bf Numerical solution with relevant disorder, $\hat V = 1$ and $N=100$}. The plots show the disorder potential $V(x)$, the induced expectation value $\langle\mathcal{O}(x)\rangle$ for the scalar field, the scalar field on the horizon $\Phi_{\mathcal H}$, and the scalar field everywhere in the spacetime.}
\end{figure}

In the numerics for this relevant case, the disorder potential will be modeled with $N = 50$ or $100$ cosines. In performing the actual numerics one can set $k_0 = 1$, which effectively sets the units. Since the Einstein-scalar equations are nonlinear, we want to make sure we can capture the first five harmonics emanating from each progenitor $k_n$ mode. In order to make sure we resolved all such scales, most of our simulations run with $100$ points in the radial direction $\tilde{y}$ and no less than $1000$ points in the Fourier direction $x$. An illustrative solution is shown in figure \ref{numerics}.
\newpage

\subsection{Thermodynamics}

This subsection will describe the dependence of the entropy density $s$ on temperature in two cases with relevant disorder ($\Delta = \frac{3}{4}$) and with a fairly strong disorder magnitude in the UV: $\hat V \equiv \bar V/\sqrt{2 \pi} = 0.5$ and $\hat V = 1$. In the case of marginal disorder, logarithmic divergences in perturbation theory could be plausibly resummed. Relevant disorder leads to power law divergences in perturbation theory that should not be naively resummed. Therefore we must turn to numerics from the outset to access the low temperature thermodynamics. In appendix \ref{sec:back} we characterize, in the spirit of \cite{Hartnoll:2014cua}, the perturbative power law divergences that appear in the averaged zero temperature metric towards the interior of the spacetime.

\begin{figure}[h]
\centering
\subfigure{\includegraphics[height = 0.2\textheight]{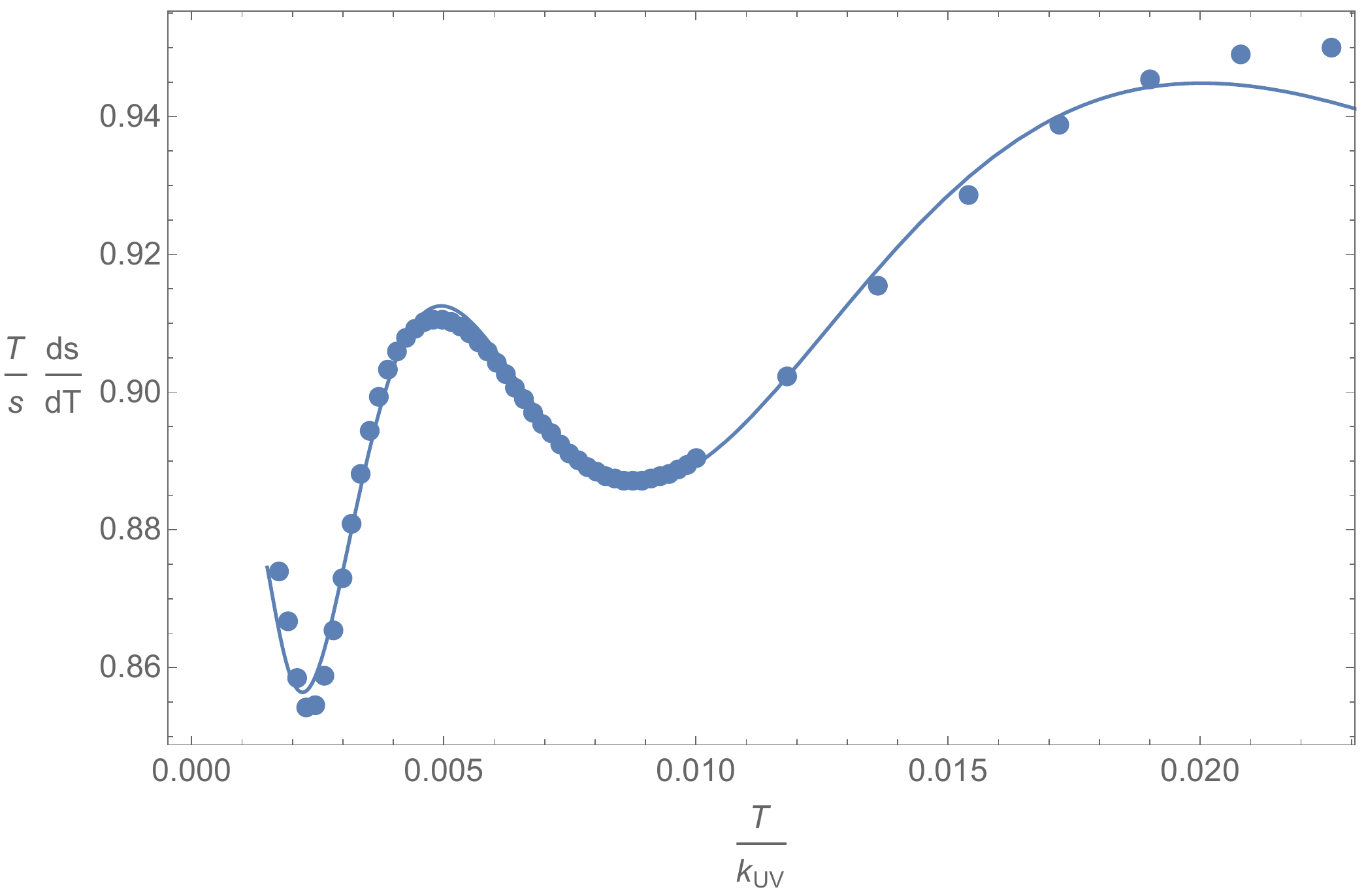}}
\subfigure{\includegraphics[height = 0.2\textheight]{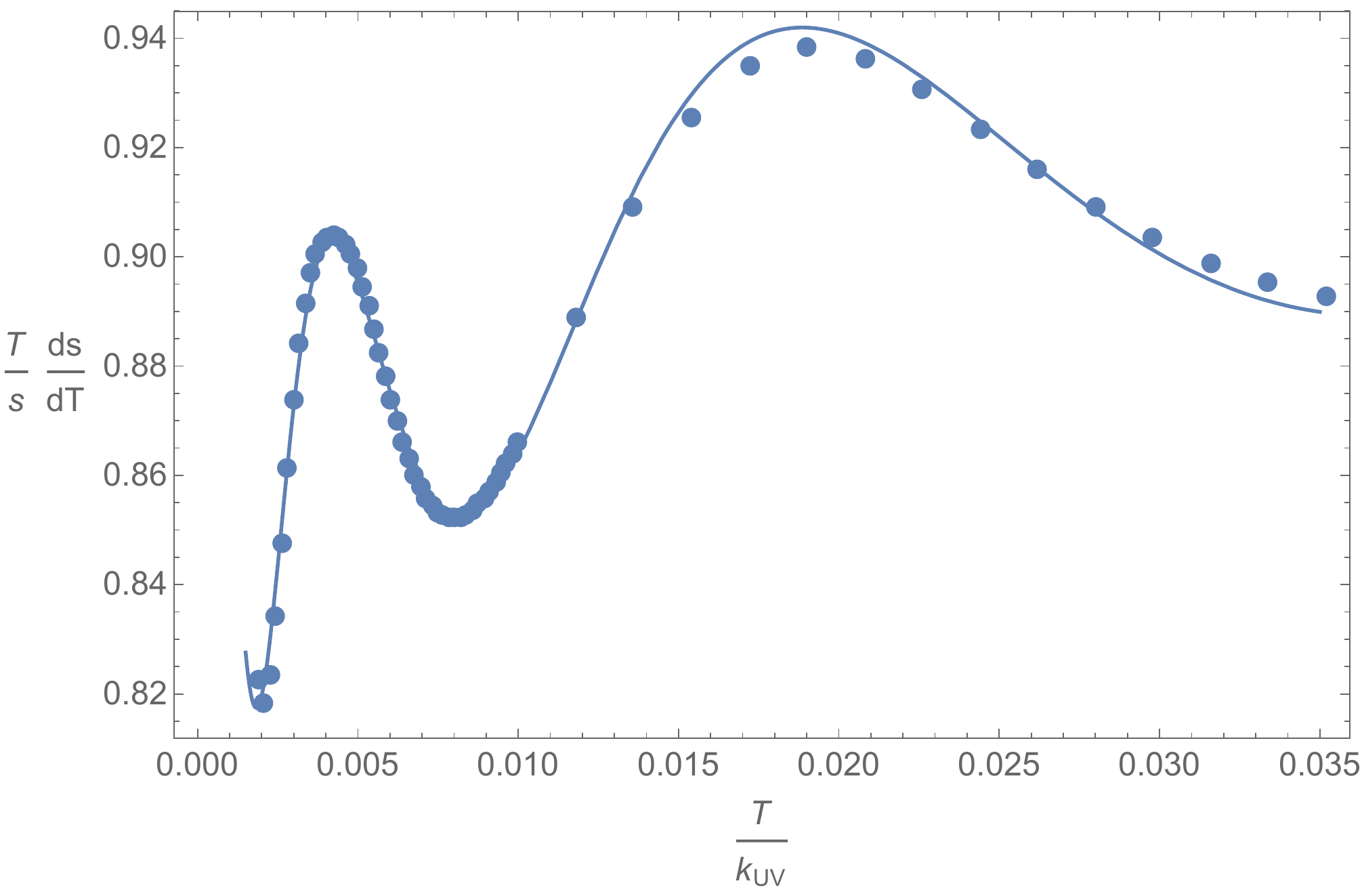}}
\caption{\label{entropy} {\bf Logarithmic derivative of the low temperature entropy density $s(T)$} for relevant disorder with $\hat V = 0.5$ (left) and $\hat V = 1$ (right). The disorder has been simulated with $N=100$ oscillatory modes. Dots are numerical data whereas the solid line is a fit to the discrete scale invariant form (\ref{eq:dsdT}). The values in the fit are $\{\gamma \approx 0.02, \delta \approx 4.5, b_0 \approx 1, b_1 \approx -0.02 \}$ for $\hat V = 0.5$ and then $\{\gamma \approx 0.03, \delta \approx 4.2, b_0 \approx 1, b_1 \approx -0.04 \}$ for $\hat V = 1$.}
\end{figure}

A plot of entropy versus temperature in these theories shows an essentially linear in temperature entropy over all temperatures. This is the result for a non-disordered one dimensional CFT, and suggests that $z$ remains equal or close to one. However, a more detailed probe at low temperatures reveals more dramatic structure. Figure \ref{entropy} shows the logarithmic derivative of the entropy with respect to temperature at low temperatures. This is the same diagnostic that was used in \cite{Hartnoll:2015faa} to conclude that $s \sim T^{1/z}$ at low temperatures in the marginal case. Figure \ref{entropy} shows that instead, in this relevant case, the numerical data is well fit by the discrete scale invariant form
\be\label{eq:dsdT}
\frac{T}{s} \frac{ds}{dT} = T^\gamma \left(b_0 + b_1 \sin \left[\delta \log \frac{T_0}{T} \right] \right) \,.
\ee
The values of the parameters appearing in the fit are given in the figure caption.
The specific heat, $c = T \, ds/dT$, is always positive in these plots, as required for thermodynamic stability.
We do not quote $T_0$, as it is ambiguous up to $T_0 \to e^{2 \pi n/\delta} T_0 $.
The precise form of this function is not the most important issue.  
Rather, what the fits in figure \ref{entropy}
show is that there is log-oscillatory structure in the thermodynamics of these models. 
The frequency of the oscillations, $\d$ in (\ref{eq:dsdT}), will be seen to be close to the frequency $\b$ of the log-oscillations
found in the thermal conductivity in the following section. The small value of $\gamma$ in (\ref{eq:dsdT}) together with $1 \approx b_0 \gg b_1$ is behind the fact that the entropy is almost linear in temperature over this range of temperatures. The nonzero $\gamma$ exponent suggests that a different temperature scaling may set in at much lower temperatures. The fact that the parameters $\g,\d,b_0$ are similar in the two cases supports the hypothesis that the solutions are flowing towards the same IR fixed point.

It seems possible that such oscillations are also present in the entropy density of the marginal case at sufficiently low temperatures.

\subsection{Thermal conductivity}

\begin{figure}[h]
\centering
\includegraphics[height = 0.3\textheight]{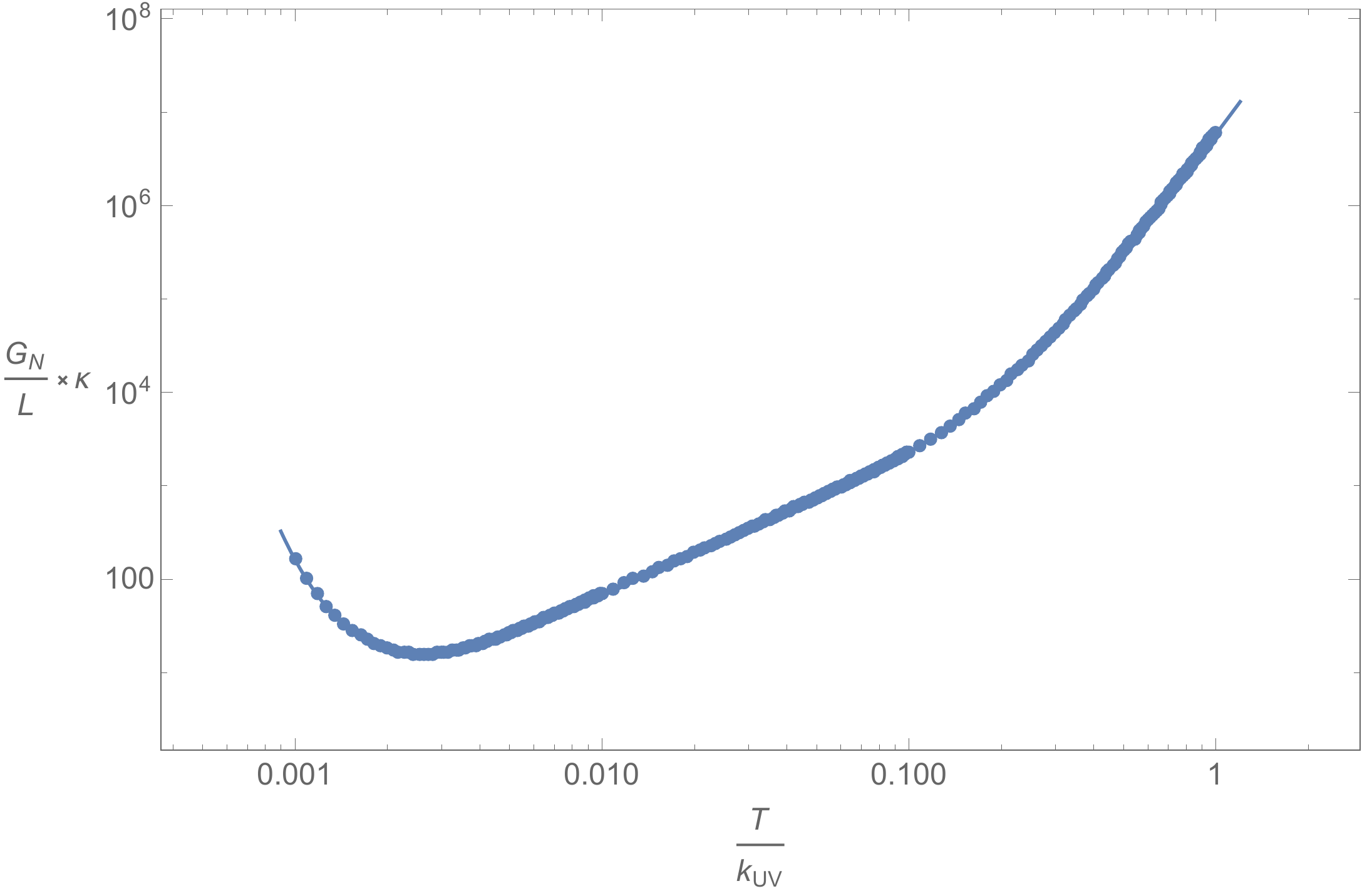}
\caption{\label{rel001}  {\bf Log-log plot of thermal conductivity $\kappa(T)$} with $\hat V \equiv \bar V/\sqrt{2 \pi} = 0.01$ and relevant disorder ($\Delta = \frac{3}{4}, d=1$). The dots are numerical data points with $N=50$ oscillator modes generating the disordered potential. The solid line is the analytic perturbative result (\ref{eq:kappapert}) and (\ref{eq:gammapert}). At high temperatures compared to the disorder cutoff $k_\text{UV}$, $\kappa \sim T^{9/2}$. In the intermediate, disordered regime $\kappa \sim T^{3/2}$. At the lowest temperatures, below the long wavelength cutoff $k_\text{IR}$, $\kappa$ increases, ultimately exponentially.}
\end{figure}

We start with a weakly disordered case to corroborate our expectations. Figure \ref{rel001} shows the thermal conductivity as a function of temperature for weak disorder, $\hat V \equiv \bar V/\sqrt{2 \pi} = 0.01$. Also shown in the same plot is the perturbative result (\ref{eq:kappapert}) and (\ref{eq:gammapert})\footnote{To capture the very low temperature regime correctly, i.e. $T \sim k_\text{IR}$, one must replace the integral in (\ref{eq:gammapert}) with a discretized sum over $N$ terms.}, now with $\Delta = \frac{3}{4}$. Once again, the numerical and perturbative results are seen to be in excellent agreement over the entire temperature range.
The log-log plot cleanly shows the three expected behaviors. At temperatures above $k_\text{UV}$, the thermal conductivity tends to $\kappa \sim T^{9/2}$, as can be seen from the analytic expressions. At temperatures below $k_\text{IR}$, the potential is effectively a lattice with wavevector $k_\text{IR}$ rather than a disordered potential. A lattice cannot efficiently relax momentum in a relativistic theory and so the conductivity is expected to increase exponentially towards low temperatures \cite{Hartnoll:2012rj}. The intermediate truly disordered temperatures show the scaling $\kappa \sim T^{3/2}$ anticipated from perturbation theory (\ref{eq:pertscaling}).

\begin{figure}[h]
\centering
\includegraphics[height = 0.3\textheight]{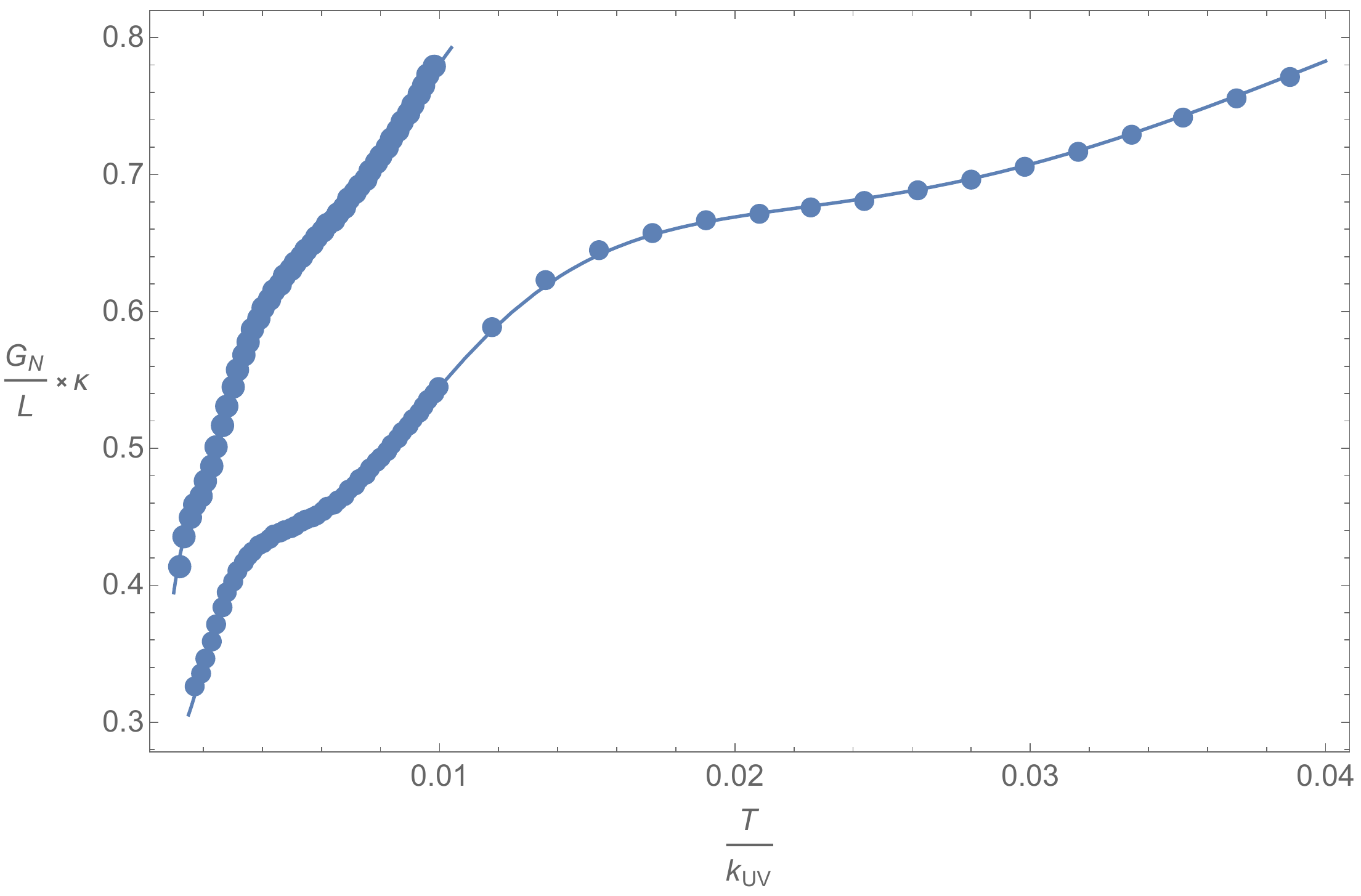}
\caption{\label{bigrel} {\bf Low temperature thermal conductivity $\kappa(T)$} for relevant disorder with $\hat V = 0.5$ (upper curve) and $\hat V = 1$ (lower curve). The disorder has been simulated with $N=100$ oscillatory modes. The solid lines are fits to the discrete scale invariant form of equation (\ref{eq:dsi}). The fit for the $\hat V = 0.5$ case has $\{\a \approx 0.3, \beta \approx 5.4, c_0 \approx 3, c_1 \approx 0.06 \}$. The fit for the $\hat V = 1$ case has $\{\a \approx 0.28, \beta \approx 4.1, c_0 \approx 2, c_1 \approx -0.09 \}$.}
\end{figure}

Figure \ref{bigrel} shows the thermal conductivity for stronger disorder, $\hat V = 0.5$ and $\hat V = 1$. The plot focusses on the disordered low temperature regime, but above the IR cutoff (these plots have used $N=100$ oscillatory modes to simulate the disordered potential, allowing us to go to lower temperatures). Also shown in the plot is a fit to the discrete scale invariant form
\be\label{eq:dsi}
\kappa_\text{d.s.i.} = T^\alpha \left(c_0 + c_1 \sin \left[\beta \log \frac{T_0}{T} \right] \right) \,.
\ee
The values of the parameters appearing in the fit are given in the figure caption.
The numerical data is seen to contain over one period of oscillation, close to two periods for the $\hat V = 1$ case. 
There are some error bars on the fits, due to the finite temperature range, but both disorder strengths are consistent with roughly $\alpha \approx 0.3$. Note that this exponent is significantly different to the perturbative exponent of $3/2$ that we saw in figure \ref{rel001}. This scaling is therefore indicative of a new strongly disordered fixed point. Also in both fits $c_1 \ll c_0$, consistent, perhaps, with the truncation to a single oscillatory frequency.
The agreement in the values of the exponent $\beta$ is less good -- this is likely because of the limited temperature range of the oscillations in the $\hat V = 0.5$ fit.

We see that although the magnitudes of the thermal conductivity are comparable to those for marginal disorder in figure \ref{margstronger} -- over this range of low temperatures --  the temperature dependence is quite different. The conductivity in the relevant case is tending to zero at low temperatures whereas the marginal case appears to be stabilized. Note that we do not expect any general lower bound on the thermal conductivity analogous to that found in \cite{Lucas:2015lna, Grozdanov:2015qia} for the electrical conductivity. This is because the thermal conductivity does not have a `pair production' term. A natural quantity to consider in incoherent transport is the diffusivity \cite{Hartnoll:2014lpa}
\be
D = \frac{\kappa}{T \, ds/dT}\,.
\ee
For the cases we have considered, the diffusivity indeed grows both towards low and high temperatures, with a minimum at intermediate temperatures.

\section{Towards possible applications}

Disorder is likely to be an important actor in many real world quantum critical systems. Some of the most interesting putative quantum critical physics in condensed matter systems -- for instance in the cuprates or heavy fermions -- involves a nonzero charge density (i.e. `metallic' quantum criticality) and is beyond the scope of the framework we have considered. Three important instances of `zero density' quantum criticality -- in which in addition disorder appears to play a key role -- are (i) Superfluid-insulator transitions \cite{SIT1,SIT2}, (ii) Metal-insulator transitions \cite{MIT1,MIT2}, (iii) Quantum hall plateaux transitions \cite{QH1,QH2}. The thermal conductivity is not straightforward to measure in these systems, but would offer direct insight into the role of disorder in the quantum critical physics.

The systems described in the previous paragraph have two spatial dimensions. While in principle the methods we have used adapt to two space dimensions, the additional computer power required, especially to get down to low temperatures with strong disorder, is substantial. The real challenge here is to obtain as explicit an analytical handle as possible on disordered horizons, beyond perturbation theory. Disordered zero temperature horizons may exist as decoupled fixed point solutions of the Einstein-scalar system in the same way that extremal horizons describe IR fixed points. Understanding such solutions on their own terms would amount to substantial progress in the theory of disordered quantum criticality.

\section*{Acknowledgements}

It is a pleasure to acknowledge interesting and relevant discussions with Maissam Barkeshli, Mike Blake, Ilya Esterlis, Steve Kivelson and Andy Lucas.
SAH is partially supported by a DOE Early Career Award and the Templeton foundation. This work was undertaken on the COSMOS Shared Memory system at DAMTP, University of Cambridge operated on behalf of the STFC DiRAC HPC Facility. This equipment is funded by BIS National E-infrastructure capital grant ST/J005673/1 and STFC grants ST/H008586/1, ST/K00333X/1.

\appendix

\section{Green's function of the scalar operator}
\label{sec:Green}

This appendix computes the low frequency Green's function for a scalar operator
in the background of a neutral black hole in $AdS_3$. We can write the background as \cite{Hartnoll:2009sz}
\be
ds^2 = \frac{L^2}{r^2} \left( - f(r) dt^2 + \frac{dr^2}{f(r)}+ dx^2 \right) \,,
\ee
with $f = 1 - r^2/r_+^2$. This spacetime is a solution to the theory (\ref{eq:action}) with no scalar
turned on. The temperature is $T = 1/(2 \pi r_+)$ and the entropy density $s = (\pi L)/(2 G_N) \times T$.

The equation of motion for the scalar field $\Phi$ about this background is
\be\label{eq:wave}
\nabla^2 \Phi = \frac{\Delta (\Delta-2)}{L^2} \Phi \,.
\ee
We have written the equation in terms of the dimension $\Delta$ of the dual operator $\ocal$.
The slick argument in \cite{Lucas:2015vna} gives the imaginary part of the retarded Green's function at low
frequencies in terms of a non-normalizable solution of (\ref{eq:wave}) with no time dependence. That is, write
$\Phi = \Phi_0(r,k) e^{i k x}$ and then \cite{Lucas:2015vna}
\be\label{eq:neat}
\lim_{\omega \to 0} \frac{\text{Im} \, G^R_{\ocal\ocal}(\omega,k)}{\omega} = \frac{L}{4 \pi G_N} \frac{1}{r_+} \Phi_0(r_+,k)^2 \,.
\ee
Here $\Phi_0(r,k)$ must be taken to be regular at the horizon and to have the non-normalizable near-boundary
behavior of $\Phi_0(r,k) \approx r^{2 - \Delta}$. The prefactor of $4 \pi G_N$ is related to the normalization of the scalar action
(\ref{eq:action}). For equation (\ref{eq:wave}), the relevant solution is
\bea
\lefteqn{\Phi_0(r,k) =  r^{2 - \Delta} {}_2F_{1} \left(1 - \frac{i k}{4 \pi T} - \frac{\Delta}{2},
1 + \frac{i k}{4 \pi T} - \frac{\Delta}{2}, 2 - \Delta , (2 \pi  T r)^2 \right)} \\
& & -  r^{\Delta} r_+^{2-2 \Delta} \frac{\Gamma(2-\Delta)}{\Gamma(\Delta)} \left|\frac{\Gamma\left(\frac{\Delta}{2} + \frac{i k}{4 \pi T} \right)}{\Gamma\left(1 - \frac{\Delta}{2} + \frac{i k}{4 \pi T}\right)}\right|^2 {}_2F_{1} \left(- \frac{i k}{4 \pi T} + \frac{\Delta}{2},
\frac{i k}{4 \pi T} + \frac{\Delta}{2}, \Delta , (2 \pi  T r)^2 \right) \,. \nonumber
\eea
Therefore from (\ref{eq:neat})
\be
\lim_{\omega \to 0} \frac{\text{Im} \, G^R_{\ocal\ocal}(\omega,k)}{\omega} = \frac{L}{4 \pi G_N} (2 \pi T)^{2 \Delta - 3} \frac{\sin^2(\pi \Delta)}{\pi^2} \Gamma(2-\Delta)^2 \left| \Gamma \left(\frac{\Delta}{2} + \frac{i k}{4 \pi T} \right) \right|^{4} \,.
\ee

\section{Horizon formula for the dc conductivity}
\label{sec:horizon}

This appendix adapts the computation in \cite{Donos:2014yya} to
obtain the thermal conductivity of Einstein-scalar theory in three bulk spacetime dimensions.

As described in the main text, the background metric takes the form
  \begin{align}
    \dd s^2 ={}& \frac{L^2}{y^2} \left[ y_+^2 \left( {-} f(y) A(y,x) \dd
        t^2 + S(y,x) \left( \dd x + F(y,x) \dd y \right)^2 \right) +
      \frac{B(y,x) \dd y^2}{f(y)} \right]\,, \label{eq:metri}
  \end{align}
  with $f = 1 - y^2$. The scalar field takes the form $\Phi = \Phi(y,x)$.
  In the main text we described the
boundary conditions satisfied by the various functions in the metric.
To obtain the thermal conductivity one can consider perturbations of
this background of the form
  \begin{align}
    \dd s^2 \to \dd s^2 + h_{ab}(y,x) \dd x^a \dd x^b + 2 t \zeta
    g_{tt} \dd t \dd x\, .\label{eq:pertmetric}
  \end{align}
  The $t$-linear source in $\delta g_{tx}$ is equivalent to a thermal gradient $\pa_x T = - T \zeta$
   \cite{Donos:2014cya}.
  
In the perturbed metric (\ref{eq:pertmetric}), construct the tensor
\be
G^{\mu \nu} = \nabla^{[\mu} k^{\nu]} \,, \qquad k = \frac{\pa}{\pa t} \,.
\ee
Note that $k$ is not a Killing vector in the perturbed spacetime.
The manipulations in \cite{Donos:2014yya}, with the additional requirement that the scalar field
be independent of $t$, show that the quantity
\be
Q \equiv \sqrt{-g} G^{xy} \,,
\ee
is independent of both $x$ and $y$. Using the metric (\ref{eq:metri})
one obtains
  \begin{align}
    Q ={}& \frac{1}{L} \frac{f^2}{2y} \frac{A^{3/2}}{B^{1/2} S^{1/2}}
    \left[ \partial_y \left( \frac{y^2 h_{tx}}{f A} \right)
      - \partial_x \left( \frac{y^2 h_{ty}}{f A} \right)
    \right]\, . \label{eq:Q}
  \end{align}
From this expression it can be verified that $\pa_x Q = \pa_y Q = 0$
using the equations of motion.

The fact that $Q$ is constant throughout space means that
it can be evaluated both on the horizon and at the boundary. Equating
the two resulting expressions will give the thermal conductivity.
Near the boundary one finds
\be\label{eq:QT}
\lim_{y \to 0} Q = 8 \pi G_N \langle T^{tx} \rangle_\text{FT}  + \ocal(t) \,.
\ee
Here $\langle T^{tx} \rangle_\text{FT}$ is the heat current (also, the momentum density) in the dual
field theory, computed as usual from the holographic stress tensor \cite{Balasubramanian:1999re}.
Note that to from the bulk stress tensor $T^{ab}$ to the field theoretic momentum density
$\langle T^{tx} \rangle_\text{FT}$, one has to multiply the bulk $T^{ab}$ by $\sqrt{\sigma} \xi_a n_b$, where $n$ is a unit
timelike vector, $\xi = \pa_x$ and the spatial measure $\sqrt{\sigma} = \sqrt{g_{xx}}$.
The time dependent part in (\ref{eq:QT}) is unimportant for our purposes \cite{Donos:2014cya, Donos:2014yya}.
  
  Now we consider the near horizon behavior of the fluctuations. We
  only need the behavior of the two metric components that appear in
  the expression (\ref{eq:Q}) for $Q$. The general behavior of these components near
  the horizon is
  \begin{align}
    h_{tx}(y,x) ={}& L y_+ \left[S^{(0)}(x)\right]^{1/2} h_{tx}^{(0)}(x)
    + h_{tx}^{(1)}(x) f(y) \log \left(\frac{1+y}{1-y} \right) +
    {\cal O}(y-1)\, , \\
    h_{ty}(y,x) ={}& h_{ty}^{(0)}(x) + {\cal
      O}(y-1)\, .
  \end{align}
  The logarithmic term is included in $h_{tx}$ so that the total
  $\delta g_{tx} = t \zeta g_{tt} + \delta h_{tx}$ can be written in
  terms of the ingoing Eddington-Finkelstein coordinate $v$, which is:
  \begin{align}
    v ={}& t - \frac{1}{2y_+} \log \left( \frac{1+y}{1-y}
    \right)\, .
  \end{align}
   Regularity (infalling boundary conditions) requires the $t$ and $y$ dependence of $\delta g_{tx}$ to combine into a dependence only on $v$. Therefore
  \begin{align}
    h_{tx}^{(1)}(x) ={}& \frac{y_+}{2} \zeta L^2 A^{(0)}(x)\, .
  \end{align}  

Plugging the near horizon behavior into $Q$, we find $Q
  = {-} y_+ h_{tx}^{(0)}(x) + {\cal O}(y-1)$. Since this must be a
  constant, we see that $\partial_x h_{tx}^{(0)} = 0$. We can go
  further and set all of the higher order terms (in $y-1$) to zero,
  i.e.~impose $\partial_y Q =0$. The leading term in this equation
  near the horizon reads, using the background equations of motion
  expanded near the horizon:
  \begin{align}
    0 ={}& y_+ \zeta + \frac{2 h_{tx}^{(0)} \left(\partial_x
        \Phi^{(0)}\right)^2}{L y_+ \sqrt{S^{(0)}}} - \partial_x \left[
      \frac{h_{ty}^{(0)}}{L^2 A^{(0)}} - \frac{h_{tx}^{(0)}}{2 L y_+
        \sqrt{S^{(0)}}} \partial_x \log A^{(0)} \right]\, .
  \end{align}
If we average this equation over a period $L_x$ in the $x$ direction,
the total derivative term drops out. This allows us to solve for
  $h_{tx}^{(0)}$. Insertion into the expression for $Q$ gives
  \begin{align}
    Q ={}& - y_+ h_{tx}^{(0)} = \frac{\zeta y_+^3}{2 X}\, , & X ={}&
    \frac{1}{L_x} \int \dd x\, \frac{(\partial_x
      \Phi^{(0)})^2}{L \sqrt{S^{(0)}}}\, . \label{eq:horizon}
  \end{align}
  Since $y_+ = 2\pi T$, the thermal conductivity is:
  \begin{align}
    \kappa ={}& \frac{\langle T^{xt}\rangle_\text{FT}}{- \pa_x T} = \frac{1}{8 \pi G_N} \frac{Q}{\zeta T} =
    \frac{\pi^2 T^2}{2 G_N} \frac{1}{X}\, . \label{eq:tomain}
  \end{align}
The second equality used the expression (\ref{eq:QT}) for $Q$ at the boundary while the third
equality used the expression (\ref{eq:horizon}) for $Q$ at the horizon. Equation (\ref{eq:tomain}) is the result for the
thermal conductivity quoted in the main text.

Let us check the perturbative limit. Here $S^{(0)} = 1$ and $\Phi^{(0)}$ is the horizon value of the solution of a linearized
scalar about the Schwarzschild-$AdS_3$ solution with the random source asymptotic boundary condition \cite{Hartnoll:2014cua, Hartnoll:2015faa}:
\be\label{eq:disorderexpand}
\Phi(r,x) = 2 \bar V \sqrt{\frac{\Delta k}{2 \pi}} \sum_{n=1}^N \Phi_0(r,k_n) \cos(k_n x + \gamma_n) \,,
\ee
where $\Delta k = k_\text{UV}/N$, $k_n = n \Delta k$, the phase $\gamma_n$ is uniformly randomly distributed between $0$ and $2 \pi$, and the solutions at each $k_n$ are normalized so that $\Phi_0 \approx r^{2 - \Delta}$ near the boundary, with unit coefficient. Therefore
\be
X = \frac{1}{L_x} \frac{1}{L} \int_0^{L_x} \left(\pa_x \Phi \right)^2 = \frac{\bar V^2}{\pi L} \sum_{n=1}^N \left(\Delta k\right) k_n^2 \Phi_0(r_+,k_n)^2 \,.
\ee
Now using (\ref{eq:neat}) and taking the continuum $\Delta k \to 0$ limit
\be
X = \frac{4 G_N \bar V^2}{L^2 T} \int_0^\infty \frac{dk}{2\pi} k^2 \lim_{\omega \to 0} \frac{\text{Im} \, G^R_{\ocal\ocal}(\omega,k)}{\omega}
= \frac{\pi T}{L} \, \Gamma \,.
\ee
Here $\Gamma$ is defined in (\ref{eq:gamma}). Hence (\ref{eq:tomain}) becomes
\be
\kappa =  \frac{s}{\Gamma} \,.
\ee
Where we used $s = (\pi L)/(2 G_N) \times T$. We have reproduced the memory matrix result (\ref{eq:kappapert}).
To get the match we have included a factor of $2\pi$ in the normalization of (\ref{eq:disorderexpand}) that was missing
in \cite{Hartnoll:2014cua, Hartnoll:2015faa}.

\section{Perturbative backreaction of relevant disorder}
\label{sec:back}

In this appendix we consider the lowest order backreaction of
relevant disorder, with $\Delta < (d+2)/2$, on the metric. We work in $d=1$ and at
zero temperature. The calculation proceeds very similarly to that in \cite{Hartnoll:2014cua},
so details will be sparse. The final result is equation (\ref{eq:resultA}), which shows that two
components of the averaged metric have power law growths as one moves towards the
interior of the spacetime. The power law growth is of course as one expects for a relevant operator.
A minor novelty compared
to the marginal case \cite{Hartnoll:2014cua} is that two rather than one metric component
diverge (in Fefferman-Graham gauge) towards the interior of the spacetime.

The scalar solution that corresponds to a
disordered source on the boundary was given in (\ref{eq:disorderexpand}).
To lowest order at zero temperature, the scalar propagates in an
AdS$_3$ background. Therefore the profile with the required
near-boundary behavior is a linear combination of Bessel functions:
\begin{align}
  \Phi(r,x) ={}& \bar V \, \frac{2^{3-\Delta} \sqrt{\frac{\Delta k}{2\pi}}}{\Gamma(\Delta - 1)} \, \sum_n r k_n^{\Delta - 1} K_{\Delta-1} (k_n \, r) \cos \left( k_n x +
    \gamma_n \right) \, , \label{eq:zero}
\end{align}
where $k_n, \gamma_n$ and $\Delta k$ are defined below (\ref{eq:disorderexpand}).
Relevant disorder for $d=1$ corresponds to $\Delta < 3/2$.

The scalar field (\ref{eq:zero}) sources metric perturbations at order $\bar V^2$,
through the Einstein equations following from the action (\ref{eq:action}). The metric to second order can be written:
\begin{align}
  \dd s^2 ={}& \frac{1}{r^2} \left[ {-} \left( 1 + \bar V^2
      A^{(2)}(r,x) \right) \dd t^2 + \dd r^2 + \left( 1 + \bar V^2
      B^{(2)}(r,x) \right) \dd x^2 \right]\, .
\end{align}
$A^{(2)}$ and $B^{(2)}$ are fixed, modulo boundary conditions, in
terms of the scalar sources via the Einstein equation. The solution
reads as:
\begin{align}
  A^{(2)}(r,x) ={}& \alpha_1(r) + \alpha_2(x) - 4 \int \dd r \dd
  x\, \partial_r \Phi \partial_x \Phi\, ,\\
  B^{(2)}(r,x) ={}& \beta_1(x) + \frac{1}{2} \beta_2(x) r^2 + 2 \int
  \dd r \, r \int \frac{\dd r}{r^3} \left[ \mu^2 \Phi^2 - r^2
    ( \partial_x \Phi)^2 - r^2 (\partial_r \Phi)^2
  \right]\, , \\
  \alpha_1(r) ={}& \eta_1 + \frac{1}{2} \eta_2 r^2 + 2\int \dd r \, r
  \int \frac{\dd r}{r^3} \left\langle\mu^2 \Phi^2 + r^2 (\partial_x
    \Phi)^2 - r^2 (\partial_r \Phi)^2 \right\rangle\, .
\end{align}
 Regularity at the Poincare
horizon $r \to \infty$ requires $\eta_2 = \alpha_2 = \beta_2 = 0$,
while $\beta_1$ is fixed by setting $A^{(2)} = B^{(2)}$ at the
conformal boundary $r \to 0$.

Upon plugging in the scalar solution (\ref{eq:zero}) and averaging over disorder
configurations, i.e.~the random phases $\gamma_n$, one can readily
extract the large $r$ behavior of the averaged metric components from
these integral expressions. We find:
\begin{align}
  \lim_{r \to \infty} \left\langle A^{(2)}(r,x) \right\rangle_R ={}&
  \eta_1 - \gamma \bar V^2 r^{3-2\Delta}\, , & \lim_{r \to \infty} \left
    \langle B^{(2)}(r,x) \right\rangle_R ={}& \beta_1(x) - \delta \bar
  V^2 r^{3-2\Delta}\, , \label{eq:resultA}
\end{align}
where $\gamma$ and $\delta$ are the positive constants
\begin{align}
  \gamma ={}& \frac{(3 - 10 \Delta + 4 \Delta^2) \Gamma \left( \Delta
      - \tfrac{3}{2} \right) \Gamma \left(2 \Delta - \tfrac{3}{2}
    \right)}{4^{\Delta} \Gamma(\Delta - 1)^2 \Gamma(\Delta + 1)}\, , &
  \delta ={}& \frac{\Gamma \left( \Delta - \tfrac{1}{2} \right) \Gamma
    \left( 2 \Delta - \tfrac{3}{2} \right)}{4^{\Delta-1}
    \Gamma(\Delta-1)^2 \Gamma(\Delta)}\, .
\end{align}
Note that as $\Delta \to \tfrac{3}{2}$, the marginal limit, $\gamma$ has a pole, and one
recovers the logarithmic divergence found and resummed in \cite{Hartnoll:2014cua}
to obtain a non-trivial dynamic critical exponent. Away from this limit, the power law
divergences cannot be resummed reliably.

\end{document}